\documentclass[11pt]{article}
\usepackage{dsfont}

\usepackage{amsmath,amssymb,graphicx}
\usepackage{diagbox}
\usepackage{hyperref}
\usepackage[nosort]{cite}

\usepackage{tikz}
\usetikzlibrary{calc} \usetikzlibrary{patterns} \usetikzlibrary{decorations.pathreplacing} \usetikzlibrary{decorations.markings} \usetikzlibrary{decorations.pathmorphing} \usetikzlibrary{positioning}

\usepackage{mathtools}

\usepackage{color}
\definecolor{darkred}{rgb}{0.65,0.15,0}
\definecolor{newgreen}{rgb}{0.2,0.62,0.14}
\hypersetup{pdfborder={0 0 0},colorlinks=true,urlcolor=blue,citecolor=blue,linkcolor=darkred,linktocpage=true}

\DeclareFontFamily{U}{mathx}{}
\DeclareFontShape{U}{mathx}{m}{n}{ <-> mathx10 }{}
\DeclareSymbolFont{mathx}{U}{mathx}{m}{n}
\DeclareFontSubstitution{U}{mathx}{m}{n}

\DeclareMathAccent{\widecheck}{0}{mathx}{"71}

\numberwithin{equation}{section}

\def\nn{\nonumber}

\def\spa#1.#2{\left\langle#1\,#2\right\rangle}
\def\spb#1.#2{\left[#1\,#2\right]}

\def\beq{\begin{equation}}
\def\eeq{\end{equation}}
\let\Re\relax
\let\Im\relax
\DeclareMathOperator{\Re}{Re}
\DeclareMathOperator{\Im}{Im}

\newcommand{\eq}{\begin{equation}}
\newcommand{\eqe}{\end{equation}}
\newcommand{\eqa}{\begin{eqnarray}}
\newcommand{\eqae}{\end{eqnarray}}

\newcommand{\bea}{\begin{eqnarray}}
\newcommand{\eea}{\end{eqnarray}}
\newcommand{\dd}{\mathrm{d}}

\newcommand{\CC}{\mathbb C}

\newcommand{\ZZ}{\mathbb Z}

\newcommand{\FFpm}[3]{{\rm F}^{\pm(#1)}_{#2,#3}}
\newcommand{\FFp}[3]{{\rm F}^{+(#1)}_{#2,#3}}
\newcommand{\FFm}[3]{{\rm F}^{-(#1)}_{#2,#3}}

\newcommand{\seedp}[3]{f^{+(#1)}_{#2,#3}}

\newcommand{\SLtwoZ}{{\rm SL}(2,\mathbb{Z})}

\newcommand{\BB}{{\rm B}}

\newbox\charbox
\newbox\slabox
\def\s#1{{      
        \setbox\charbox=\hbox{$#1$}
        \setbox\slabox=\hbox{$/$}
        \dimen\charbox=\ht\slabox
        \advance\dimen\charbox by -\dp\slabox
        \advance\dimen\charbox by -\ht\charbox
        \advance\dimen\charbox by \dp\charbox
        \divide\dimen\charbox by 2
        \raise-\dimen\charbox\hbox to \wd\charbox{\hss/\hss}
        \llap{$#1$}
}}

\usepackage[shadow,textwidth=2.7cm]{todonotes}
\usepackage{ifthen}
\reversemarginpar
\newcounter{todocounter}

\colorlet{ddcolor}{green!40!white}

\newcommand{\ddinline}[2][]{
  \ifthenelse { \equal {#1} {} }
    { \def\temp {#2} }  
    { \def\temp {#1} }   
  \refstepcounter{todocounter}\todo[color=ddcolor,inline,caption={\textbf{\thetodocounter. DD} \temp}]{\textbf{\thetodocounter. DD:} #2}{}}

\colorlet{rtcolor}{blue!20!white}

\newcommand{\rtinline}[2][]{
  \ifthenelse { \equal {#1} {} }
    { \def\temp {#2} }  
    { \def\temp {#1} }   
  \refstepcounter{todocounter}\todo[color=rtcolor,inline,caption={\textbf{\thetodocounter. RT} \temp}]{\textbf{\thetodocounter. RT:} #2}{}}

\colorlet{akcolor}{yellow!40!white}

\newcommand{\akinline}[2][]{
  \ifthenelse { \equal {#1} {} }
    { \def\temp {#2} }  
    { \def\temp {#1} }   
  \refstepcounter{todocounter}\todo[color=akcolor,inline,caption={\textbf{\thetodocounter. AK} \temp}]{\textbf{\thetodocounter. AK:} #2}{}}

\newcommand{\GG}{ {\rm G} }
\newcommand{\EE}{ {\rm E} }

\newcommand{\PS}{\sum_{\gamma \in B(\ZZ)\backslash {\rm SL}(2,\ZZ)}}
\newcommand{\seeed}{\varphi}
\newcommand{\summ}{\Phi}

\begin{document}

\mbox{ }
\vspace{10mm}

\begin{center}

{\bf {\LARGE \sc To the cusp and back: \\[2mm]
Resurgent analysis for modular graph functions }}

\vspace{10mm}
\normalsize
{\large  Daniele Dorigoni${}^{1}$, Axel Kleinschmidt${}^{2,3}$ and Rudolfs Treilis${}^1$}

\vspace{10mm}
${}^1${\it Centre for Particle Theory \& Department of Mathematical Sciences\\
Durham University, Lower Mountjoy, Stockton Road, Durham DH1 3LE, UK}
\vskip 1 em
${}^2${\it Max-Planck-Institut f\"{u}r Gravitationsphysik (Albert-Einstein-Institut)\\
Am M\"{u}hlenberg 1, DE-14476 Potsdam, Germany}
\vskip 1 em
${}^3${\it International Solvay Institutes\\
ULB-Campus Plaine CP231, BE-1050 Brussels, Belgium}

\vspace{20mm}

\hrule

\vspace{5mm}
\begin{tabular}{p{14cm}}
Modular graph functions arise in the calculation of the low-energy expansion of closed-string scattering amplitudes. For toroidal world-sheets, they are $\SLtwoZ$-invariant functions of the torus complex structure that have to be integrated over the moduli space of inequivalent tori. We use methods from resurgent analysis to construct the non-perturbative corrections arising when the argument of the modular graph function approaches the cusp on this moduli space. $\SLtwoZ$-invariance will in turn strongly constrain the behaviour of the non-perturbative sector when expanded at the origin of the moduli space.
\end{tabular}

\vspace{6mm}
\hrule
\end{center}

\thispagestyle{empty}

\newpage
\setcounter{page}{1}

\setcounter{tocdepth}{2}
\tableofcontents

\bigskip

\section{Introduction}
\label{sec:1}

In recent years, there has been considerable progress in understanding the low-energy expansion of string scattering amplitudes.
Of particular importance to both the physics and mathematics communities is the study of the low-energy expansion for closed-string genus one amplitudes, i.e. for strings whose world-sheet is a two-dimensional torus. This problem led to the introduction of an infinite class of
non-holomorphic objects: the so-called {\it modular graph forms} (MGFs) \cite{Green:1999pv,Green:2008uj,DHoker:2015gmr, DHoker:2015wxz, DHoker:2016mwo}. These functions depend on a parameter $\tau$ in the Poincar\'e upper half-plane associated with the moduli space of the world-sheet torus on which the modular group $\SLtwoZ$ acts in the standard fashion. MGFs are characterised by having good transformation properties under the modular group, and they are constructed from world-sheet Feynman-type diagrams. However, they are not necessarily holomorphic functions of $\tau$.

Amongst the many remarkable properties of these objects we have: the appearance of multiple zeta values in their expansion around 
the cusp $\tau \rightarrow i \infty$, with $\tau$ the modular parameter of the torus, and an intricate network of algebraic and differential relations. The study of these properties and their implications has received attention in both the physics
\cite{Green:1999pv, Green:2008uj,DHoker:2015gmr , DHoker:2015wxz, DHoker:2016mwo, Green:2013bza,  DHoker:2015sve, Basu:2015ayg, Basu:2016xrt, Basu:2016kli, Basu:2016mmk, DHoker:2016quv, Kleinschmidt:2017ege, Basu:2017nhs, Broedel:2018izr, Ahlen:2018wng,Gerken:2018zcy, Gerken:2018jrq, DHoker:2019txf, Dorigoni:2019yoq, DHoker:2019xef, DHoker:2019mib, DHoker:2019blr, Basu:2019idd,  Gerken:2019cxz, Hohenegger:2019tii, Gerken:2020yii, Basu:2020kka, Vanhove:2020qtt, Basu:2020pey, Basu:2020iok , Hohenegger:2020slq,Dorigoni:2021jfr,Dorigoni:2021ngn}
and mathematics literature \cite{Brown:mmv, Zerbini:2015rss, Brown:I, Brown:II, DHoker:2017zhq,Zerbini:2018sox, Zerbini:2018hgs, Zagier:2019eus, Berg:2019jhh,Drewitt:2021}. A review and a {\tt Mathematica} implementation can be found in \cite{Gerken:review,Gerken:2020aju}, see also \cite{DHoker:2022dxx} for a recent comprehensive review. 

The study of these modular objects has been tackled from multiple directions.
In particular, one can try to directly evaluate the world-sheet integrals in closed-string genus-one amplitudes thus obtaining lattice-sum representations of MGFs \cite{Green:1999pv, Green:2008uj, DHoker:2015gmr, DHoker:2015wxz, Gerken:2018jrq}. 
Although it is possible to extract the asymptotic expansion at the cusp $\tau \rightarrow i \infty$ from some of these lattice-sum representations \cite{Zagier:notes, DHoker:2017zhq}, this is nonetheless a hard task suggesting that a different approach might in general be necessary.

Crucially, the differential structure~\cite{DHoker:2016mwo}  satisfied by the MGFs suggests a different approach. The intricate differential relations amongst the various MGFs can be made manifest by representing them via iterated integrals over holomorphic Eisenstein series ${\rm G}_k(\tau)$ and their complex conjugates \cite{DHoker:2015wxz, Broedel:2018izr, Gerken:2020yii}. 
While the lattice-sum representations mentioned above manifest the modular properties of these objects and lead to the interpretation of MGFs as discretised Feynman integrals for a scalar field on the torus, we have in turn that the iterated-Eisenstein-integral representations expose the full structure of algebraic and differential relations of MGFs, making their Fourier-expansion, and hence the asymptotic expansion at the cusp $\tau \rightarrow i \infty$, amenable to study.
 
Recently, a combined approach was introduced by presenting Poincar\'e-series representations of MGFs restricted to the modular-invariant case, i.e.\ modular graph functions and modular-invariant combinations of forms~\cite{Dorigoni:2021jfr,Dorigoni:2021ngn}. Poincar\'e series are an extremely convenient way of rewriting a modular-invariant function as a sum over images under the modular group $\SLtwoZ$ of a simpler function that is usually called its (Poincar\'e) seed function. 

For illustration we can consider MGFs associated with one-loop graphs with $k\geq 2$ links. 
These modular-invariant objects are known to be given by non-holomorphic Eisenstein series $\EE_k(\tau)$, which can be expressed as very simple iterated integrals of a single holomorphic Eisenstein series, hence they all are said to be of depth one. From the Poincar\'e-series point of view, it is known that the non-holomorphic Eisenstein series $\EE_k(\tau)$ can be written as a sum over $\SLtwoZ$ images of the simple monomial seed $(\Im \tau)^k$ which can be assigned depth zero, see e.g.~\cite{Iwaniec:2002,Fleig:2015vky}. Using a Poincar\'e-series representation has reduced the depth (hence in general the complexity) of the functions at play.

In \cite{Dorigoni:2021jfr,Dorigoni:2021ngn}, the dictionary between lattice sums and iterated Eisenstein integrals was advanced
to the depth-two case, i.e.\ to iterated integrals of two holomorphic Eisenstein series.
We should stress that the notion of depth for an MGF is in general different from the loop order of the graph defining it.
While, as we just remarked, MGFs corresponding to one-loop Feynman graphs
can indeed be represented by iterated Eisenstein integrals of depth one,
the same is not generically true from two loops onwards. 
In particular, two-loop MGFs do not exhaust all depth-two modular invariant objects and there are infinitely many two-loop MGFs that can be reduced to one-loop ones and
odd zeta values \cite{Green:2008uj, DHoker:2015gmr}. The notion of depth and loop order do not always agree, but depth is bounded from above by the loop order of the world-sheet diagram.

Among the real MGFs of depth two, the most prominent instances are the two-loop 
lattice sums $C_{a,b,c}(\tau)$ \cite{DHoker:2015gmr} built from integrals over $w{=}a{+}b{+}c$ closed-string 
Green functions. Their asymptotic expansion at the cusp $\tau \rightarrow i \infty$ can be written in terms of a truncating Laurent polynomial in $y= \pi \,\mbox{Im}\,\tau$ with rational coefficients multiplying at most a bilinear in odd zeta values~\cite{DHoker:2017zhq}, as well as carrying uniform transcendentality $w$ in a sense that we will make precise later on.

To make the story clearer, if we perform a Fourier-expansion in $\mbox{Re}\,\tau$, we have that the asymptotic expansion of the zero Fourier mode at the cusp $\tau \rightarrow i \infty$
contains the aforementioned truncating Laurent polynomial, plus an infinite tower of exponentially suppressed corrections of the form $(q\bar q)^n = \exp(-4\pi n\,\mbox{Im}\,\tau)$ with $q=\exp(2\pi i \tau)$ and $n$ a strictly positive integer.
Although these terms are negligible at the cusp, they are of fundamental importance in understanding the $\tau\to0$ limit since they produce singular contributions in this regime which will cancel some of the singular terms coming from the Laurent polynomial \cite{Green:2014yxa}.

In this paper we will combine the Poincar\'e-series representations introduced in \cite{Dorigoni:2021jfr,Dorigoni:2021ngn} with resurgent analysis, see \cite{Dorigoni:2014hea} for a recent introduction, along the lines of previous works \cite{Dorigoni:2019yoq,Dorigoni:2020oon}. Firstly we will suitably deform the Poincar\'e seed relevant for all depth-two MGFs and consider their modified behaviours at the cusp. Contrary to the undeformed case, where the asymptotic expansion truncates after finitely many terms, we will see that the deformed seed gives rise to an asymptotic, factorially divergent perturbative tail.
Using resurgence theory we will produce a non-perturbative resummation of these asymptotic series which, upon sending the deformation parameter to zero, will reproduce both the truncating perturbative Laurent polynomial at the cusp, as well as the correct tower of non-perturbative, exponentially suppressed $(q\bar{q})$-terms, in a nice example of what is usually called Cheshire Cat resurgence \cite{Dunne:2016jsr,Kozcaz:2016wvy,Dorigoni:2017smz,Dorigoni:2019kux}. 
The general functional form of the exponentially suppressed terms for the $C_{a,b,c}$ has been determined in~\cite{DHoker:2017zhq} based on the structure of the Laplace equation satisfied by them, see section~\ref{sec:2.1} below.

Furthermore, since $\tau\to i\infty$ and $\tau\to0$ are related by the usual $\SLtwoZ$ S-duality transformation $\tau\to - 1/ \tau$, we have that modularity strongly intertwines the small-$y$ behaviour of the infinite tower of $(q\bar{q})^n$ terms, no longer exponentially suppressed in this regime, with the Laurent polynomial part of the MGFs. If resurgent analysis allows us to retrieve the exponentially suppressed $(q\bar{q})^n$ corrections from perturbative data at the cusp $\tau\to i\infty$, modularity will make it possible to extract the Laurent polynomial from the  $(q\bar{q})^n$ sector at the origin $\tau\to0$.

\medskip

This paper is structured as followed. We review some of the basic definitions for MGFs and Poincar\'e series in section~\ref{sec:2}. This also includes the definition of a certain space of depth-two functions that transcends the space of MGFs and is easier to analyse. In section~\ref{sec:3}, we then apply methods from resurgent analysis to MGFs to derive the exponentially suppressed terms (for $\tau\to i \infty$) in the zero Fourier modes of the MGFs. We present several examples and also combine the analysis with modular invariance to study the analogue of the strong coupling behaviour $\tau \to 0$ in section~\ref{sec:4}. In section~\ref{sec:Conc}, we offer some concluding remarks. Two appendices contain additional technical details.

\section{Review }
\label{sec:2}

In this section we recall some fundamental features of MGFs such as their Poincar\'e series, Fourier series decomposition and expansion around the cusp $\tau\to i\infty$. We also review some of the necessary results of \cite{Dorigoni:2021jfr,Dorigoni:2021ngn} and explain how they generalise the space of MGFs.

\subsection{Modular graph functions}
\label{sec:2.1}

String perturbation theory is naturally organised into different topological sectors. For closed-string amplitudes at genus one the world-sheet is a torus  ${\mathfrak T} = \mathbb C/(\mathbb Z + \tau \mathbb Z)$ with complex modular parameter lying in the upper complex half-plane $\tau\in\mathcal{H}=\{\tau\in\mathbb{C}:\Im\tau>0\}$. In order to calculate the amplitude of a scattering process, one introduces punctures $z_j\in\mathfrak T$ and integrates them over all inequivalent configurations. Hence, in considerations of the kinematic part of the amplitude, one encounters expressions like \cite{Green:1999pv, Green:2008uj}: 
\beq
{\cal M}_n(s_{ij},\tau) = \bigg( \prod_{j=2}^n \int_{\mathfrak T} \frac{\dd^2 z_j}{\Im \tau} \bigg) \exp\bigg( \sum_{1\leq i<j}^n s_{ij} G(z_i{-}z_j,\tau) \bigg)\,,
\label{rev.01}
\eeq
where translational invariance can be used to set $z_1$ to an arbitrary value. In this expression $s_{ij}\in\mathbb{C}$ are dimensionless Mandelstam invariants, which we take to be independent complex numbers, and $G(z,\tau)$ is the Green function on a torus given by 
\beq
G(z,\tau) = \frac{ \Im \tau}{\pi} \sum_{(m,n) \neq (0,0)} \frac{ e^{2\pi i (mv-nu)}}{|m\tau {+}n|^2}\,,
\label{rev.02}
\eeq
where $z = u\tau +v$ with $u,v\in [0,1)$.
This sum is only conditionally convergent and is understood using the Eisenstein summation convention~\cite{ApostolTomM1976MfaD}.

To proceed, we remind the reader that there is a natural action of $\SLtwoZ$ --~the modular group\footnote{Only ${ \rm PSL} (2,\mathbb{Z})=\SLtwoZ/\{\pm 1\}$ acts faithfully on the upper half-plane, but this distinction plays no role for us.}~-- on $\tau\in\mathcal{H}$, which is given by
\begin{equation}
    \label{eq_mod_trans}
    \gamma =
    \begin{pmatrix}
    a & b\\
    c & d
    \end{pmatrix}
    \in \SLtwoZ\,,
    \qquad
    \gamma\cdot\tau = \frac{a\tau + b}{c\tau + d}.
\end{equation}
The string amplitude is required to be invariant under modular transformations. This is clear from (\ref{rev.01}), since both the integrand as well as the measure are invariant under transformation (\ref{eq_mod_trans}). It will also be useful to consider the Borel subgroup 
$B(\mathbb{Z}) = \big\{\big(\begin{smallmatrix}
  \pm1 & n\\
  0 & \pm1
\end{smallmatrix}\big) : n\in\mathbb{Z}\big\}$
, which corresponds to translations $\tau\to\tau+n$.

When one Taylor expands the exponential in (\ref{rev.01}) in the $s_{ij}$, one is naturally led to a graphical scheme for organising the terms that emerge - these objects are called {\it modular graph functions} (MGFs) and were introduced in \cite{DHoker:2015wxz}. In order to construct a graph out of the terms in the series, we associate a vertex with each of the punctures $z_1,z_2,...,z_n$ and an edge connecting vertices $i$ and $j$ with each occurrence of the propagator $G(z_i-z_j,\tau)$. In turn, every vacuum graph produced from a scalar field theory defined on a torus will also be associated to a modular graph function. We  define the \textit{weight} of an MGF as the number of edges in the corresponding graph (which is also the number of Green functions in the chosen monomial). It is important to note that weight as defined here is distinct from modular weight, which is vanishing for all MGFs.

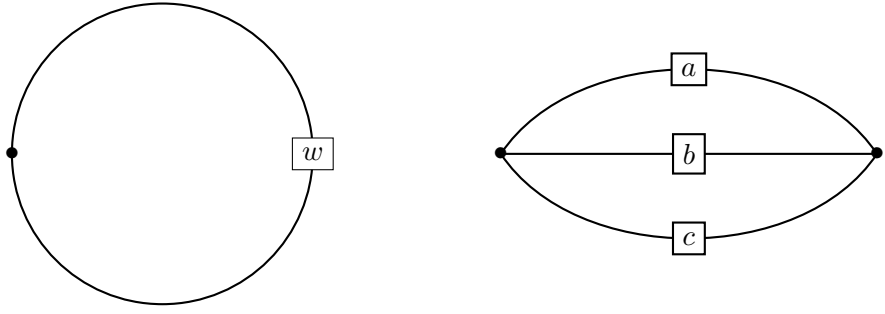
\begin{figure}[t!]
\centering
\begin{tikzpicture}
\draw [thick] (-2,0) circle (2);
\draw (-4,0)node{$\bullet$};
\draw (0,0) node [rectangle,fill=white,draw=black] {$w$};
\scope[xshift=1.5cm]
    \draw[thick,postaction={decorate}] (1,0)node{$\bullet$} .. controls (2,1.5) and (5,1.5) .. node[fill=white,draw=black]{$a$} (6,0)node{$\bullet$};
    \draw[thick,postaction={decorate}] (1,0) .. controls (2,0) and (5,0) .. node[fill=white,draw=black]{$b$} (6,0);
    \draw[thick,postaction={decorate}] (1,0) .. controls (2,-1.5) and (5,-1.5) .. node[fill=white,draw=black]{$c$} (6,0);
    \endscope
\end{tikzpicture}
\caption{\label{fig:12}\textit{The graphs corresponding to the one-loop and two-loop modular graph functions $\EE_w$ and $C_{a,b,c}$ where a link with a boxed number $w$ indicates $w$ concatenated Green functions.}}
\end{figure}

In order to understand the structure of MGFs a little better, it is useful to parametrise the punctures as $z_j = u_j \tau + v_j$ with $u_j,v_j \in [0,1)$
and $\frac{\dd^2 z_j}{\Im \tau} = \dd u_j \, \dd v_j$. In this case we use the lattice-sum representation of the Green function (\ref{rev.02}) to observe that each integral over a puncture simply enforces momentum conservation at the associated vertex. Since the torus is a compact space, the momenta are discrete and form a two-dimensional lattice (with origin removed)
\beq
p= m \tau + n \in \Lambda' \, , \ \ \ \ \ \ \Lambda' =( \mathbb Z + \tau \mathbb Z) \setminus \{0\}.
\label{rev.03}
\eeq
We are thus guaranteed that every one-particle reducible graph vanishes, since the momentum flowing through the reducible edge must be 0. As a result, the simplest non-trivial MGFs appear at one loop and are non-holomorphic Eisenstein series of weight $w > 1$
\beq
{\rm E}_w(\tau) = \bigg( \frac{ \Im \tau}{\pi} \bigg)^w \sum_{p \in \Lambda'} \frac{1}{|p|^{2w}}\,.
\label{rev.04}
\eeq
At two loops, every connected MGF can be expressed as a function $C_{a,b,c}$ of weight $w=a{+}b{+}c$:
\beq
C_{a,b,c}(\tau) =  \bigg( \frac{ \Im \tau}{\pi} \bigg)^{a+b+c} \sum_{p_1,p_2,p_3 \in \Lambda'}
\frac{\delta(p_1{+}p_2{+}p_3)}{|p_1|^{2a}|p_2|^{2b}|p_3|^{2c}} \, .
\label{rev.05}
\eeq
The graphs corresponding to the MGFs $\EE_w$ and $C_{a,b,c}$ are depicted in figure~\ref{fig:12}. There are obvious ways how one may construct MGFs at higher loop order \cite{DHoker:2015wxz} or even generalise to objects that carry non-zero modular weight, so called modular graph \textit{forms} \cite{DHoker:2016mwo}, but in this paper we only analyse the two-loop, modular-invariant case.

MGFs have a variety of interesting connections to number theory. For example, when computing the asymptotic expansion near the cusp $\tau\to i \infty$ of MGFs it is natural to encounter multiple zeta values (MZVs), i.e. generalisations of the Riemann zeta function defined by iterated (conical) sums.
The weight of an MGF can be identified with the transcendental weight of the corresponding MZV\footnote{Although not of crucial importance for this work, MZVs are defined by the conical sum $
\zeta_{n_1,n_2,\ldots,n_r} = \sum_{0< k_1 <k_2<\ldots <k_r} k_1^{-n_1}k_2^{-n_2}\ldots k_r^{-n_r}$ with $n_i \in \mathbb{N}$ and $n_r \geq 2$. The transcendental weight of a MZV is given by $w= \sum_{1\leq i \leq r }n_i$ while its depth by $r$.}  \cite{Brown:2013gia} appearing in the expansion at the cusp, however in general the loop order of an MGF is only an upper bound on the maximal depth of its possible MZVs.

Furthermore, one finds that there exists an intricate web of connections between different MGFs and $\mathbb Q$-linear combinations of multiple zeta values. Some easy examples were first discussed in \cite{DHoker:2015gmr}:
\beq
C_{1,1,1}(\tau) = {\rm E}_3(\tau) + \zeta_3 \, , \ \ \ \ \ \
C_{2,2,1}(\tau) = \frac{2}{5} {\rm E}_5(\tau) + \frac{ \zeta_5}{30} \, .
\label{rev.07}
\eeq
Observe that both sides of equations (\ref{rev.07}) are consistent with the defined weight assignments if furthermore the Riemann zeta\footnote{In this paper, we shall write the Riemann zeta function either as $\zeta(s)$ or as $\zeta_s$ depending respectively on whether $s$ is generic or fixed to some specific integer value.} $\zeta_w$ is assigned weight $w$.

 On the other hand the loop order is not preserved, since a two-loop function is expressed in terms of lower-loop objects. This is a generic feature: algebraic relations between MGFs respect the weight grading but mix different loop orders (i.e. the loop order is only a filtration). In section \ref{sec:laplace_system} we introduce the notion of depth as a more useful alternative to loop order, at least for classifying algebraic relations.

Additionally to the algebraic relations discussed before, there are also differential equations relating different MGFs. Since modular functions are naturally defined in the hyperbolic plane, the equations they satisfy are with respect to the ${\rm{SL} }(2,\mathbb{R})$ invariant Laplacian $\Delta = 4(\Im\tau)^2\partial_\tau\partial_{\bar{\tau}}$. At two loops it can be shown \cite{DHoker:2015gmr} that
\begin{align}\label{eq:diff_Cabc}
\Delta C_{a,b,c} &=(a(a-1)+b(b-1)+c(c-1))C_{a,b,c}\\
&\quad +ab(C_{a-1,b+1,c}+C_{a+1,b-1,c}+C_{a+1,b+1,c-2}-2C_{a,b+1,c-1}-2C_{a+1,b,c-1})\notag\\
&\quad+bc(C_{a,b-1,c+1}+C_{a,b+1,c-1}+C_{a-2,b+1,c+1}-2C_{a-1,b,c+1}-2C_{a-1,b+1,c})\notag\\
&\quad+ca(C_{a+1,b,c-1}+C_{a-1,b,c+1}+C_{a+1,b-2,c+1}-2C_{a+1,b-1,c}-2C_{a,b-1,c+1}),\notag
\end{align}
where one of the indices on the right hand side might get reduced to $0$ or $-1$, in which case the two-loop function is replaced by
\beq
C_{w-\ell,\ell,0}=\EE_\ell \EE_{w-\ell}-\EE_w \,  , \qquad C_{w+1-\ell,\ell,-1}=\EE_\ell \EE_{w-\ell}+\EE_{\ell-1}\EE_{w-\ell+1}.
\eeq
Formally in this procedure the divergent non-holomorphic Eisenstein series  $\EE_1$ can appear, but it always cancels out of the final answer. It was shown in \cite{DHoker:2015gmr} that the system of linear equations (\ref{eq:diff_Cabc}) can be diagonalised by the introduction of eigenfunctions $\mathcal{C}_{w;m;p}$, which are linear combinations of different $C_{a,b,c}$ with a fixed weight $w=a+b+c$. These eigenfunctions then satisfy a significantly more manageable differential equation
\beq
(\Delta -(w-2m)(w-2m-1))\mathcal{C}_{w;m;p} = t^{(0)}_{w;m;p}\EE_w+\sum_{\ell=2}^{[w/2]}t^{(\ell)}_{w;m;p}\EE_\ell \EE_{w-\ell},
\label{eq:diff_Cwmp}
\eeq
where $t^{(0)}_{w;m;p}$ and $t^{(\ell)}_{w;m;p}$ are constants, $m$ is a label for the eigenvalue of the differential equation, and $p$ labels the degeneracy of the fixed eigenspace. The explicit coefficients connecting the two bases $C_{a,b,c}$ and $\mathcal{C}_{w;m;p}$ can be found in \cite{DHoker:2021ous}.

While the representation in terms of lattice sums is convenient for a graphical interpretation and establishing connections to MZVs, the sums are hard to manipulate and many identities are hidden. Moreover, we will be interested in finding the asymptotic behaviour of MGFs as $\tau\to i\infty$, which is a task of considerable difficulty from the perspective of lattice sums. Instead it is much more convenient to use the differential equations satisfied by MGFs such as \eqref{eq:diff_Cabc} and \eqref{eq:diff_Cwmp} and solve them using a Poincar\'e series ansatz. We shall discuss said method in the following sections.
Due to the $\SLtwoZ$ invariance all modular graph functions have period one in the real part of $\tau$, hence they may be Fourier expanded in $\Re \tau$. This expansion contains a lot of information about the behaviour of the function as the modular parameter approaches the cusp $\tau\to i\infty$. To proceed, we introduce the following variables
\beq
y  = \pi \Im \tau \, , \ \ \    \ \ \ q = e^{2\pi i \tau}\, , \ \ \    \ \ \ \bar q = e^{-2\pi i  \bar \tau} \, ,
\label{rev.08}
\eeq
in which the non-holomorphic Eisenstein series (\ref{rev.04}) for positive integer weight $w>1$ can be written as the Fourier series
\begin{align}
\label{eq:FEk}
\EE_w(\tau) &=(-1)^{w-1} \frac{ {\rm B}_{2w} }{(2w)!} (4y)^w + 
\frac{4(2w{-}3)!  \zeta_{2w-1}}{(w{-}2)!(w{-}1)!} (4y)^{1-w}
\\
&\quad + \frac{2}{\Gamma(w)} \sum_{n=1}^{\infty} n^{w-1} \sigma_{1-2w}(n) \left[\sum_{a=0}^{w-1} (4ny)^{-a} \frac{\Gamma(w{+}a)}{a!\, \Gamma(w{-}a)}\right] (q^n+\bar{q}^n),
\notag
\end{align}
where $\sigma_{s}(n) = \sum_{d | n} d^s$ is a divisor sum and we have introduced the Bernoulli numbers ${\rm B}_{2w}$ that are rational numbers related to even Riemann zeta values by
\begin{equation}
\label{eq:zB}
2\zeta_{2w} = (-1)^{w+1} \frac{(2\pi)^{2w}}{(2w)!} {\rm B}_{2w}\,, \quad\quad w=1,2,3,\ldots\,.
\end{equation}

The general Fourier expansion for a modular graph function is quite similar to equation (\ref{eq:FEk}) --- one can show that MGFs grow at most polynomially at the cusp, and the expansion must be of the form $\sum_{M,N=0}^\infty L_{M,N}(y)q^M\bar{q}^N$ with $L_{M,N}(y)$ a Laurent polynomial \cite{DHoker:2015wxz}. The dominant behaviour at the cusp clearly comes from $L_{0,0}$. Some examples at two-loop level are as follows \cite{Green:2008uj, DHoker:2015gmr}

\begin{align}
C_{2,1,1}(\tau) &= \frac{ 2y^4 }{14175} + \frac{ \zeta_3 y}{45} + \frac{5 \zeta_5}{12y} - \frac{ \zeta_3^2 }{4y^2}+ \frac{ 9 \zeta_7}{16y^3} +O(q,\bar q)\,,
\label{rev.10}
\\
C_{2,2,2}(\tau) &= \frac{38y^6}{91216125} + \frac{\zeta_7}{24y} - \frac{7\zeta_9}{16y^3} + \frac{15\zeta_5^2}{16y^4}
- \frac{81\zeta_{11} }{128y^5} +O(q,\bar q)\, .
\notag
\end{align}

Notice the recurrent appearance of odd zeta values in the expansion, as well as uniform transcendental weight $w=a+b+c$ for each term once an assignment of weight 1 is given to $y=\pi \Im \tau$. Unlike the case of Eisenstein series, the zero Fourier mode gets additional, exponentially suppressed contributions from the terms  $L_{N,N}(q\bar{q})^N$ with $N>0$. 

The focus of the present work is precisely to reconstruct the $(q\bar{q})^N$ non-perturbative terms to the zero Fourier mode from the purely perturbative Laurent polynomials, or rather a suitable deformation thereof, using resurgent analysis following similar methods to the ones developed in \cite{Dorigoni:2019yoq,Dorigoni:2020oon}.
The structure of the differential equation~\eqref{eq:diff_Cwmp} fixes the functional form of these exponentially suppressed terms (see~\cite[Thm.~1.3]{DHoker:2017zhq}) in terms of incomplete Gamma functions and Laurent polynomials. We shall not rely on these results and arrive at fully explicit expression from resurgent analysis. 
%

\subsection{Depth-two Laplace systems}\label{sec:laplace_system}

One of the key results of \cite{Dorigoni:2021jfr,Dorigoni:2021ngn} was to show that all two-loop modular invariant graph functions can be written as rational linear combinations of real and imaginary modular invariant functions denoted by $\FFp{s}{m}{k}$ and $\FFm{s}{m}{k}$, respectively, and labelled by positive integers $s,m,k$ where the $\pm$ signifies that they are even (odd) under the involution $\tau \rightarrow - \bar \tau$ of the upper half-plane.

These modular invariant functions $\FFpm{s}{m}{k}$ do determine all MGFs of depth two, possibly by adding single Eisenstein series or single odd zeta values, but actually provide a wider class of modular invariant objects compared to MGFs. In particular, not all $\FFpm{s}{m}{k}$ can be expressed as lattice sums, hence they necessarily transcend the realm of MGFs, however they are still expressible in terms of iterated integrals over holomorphic modular forms (including cuspidal ones) of depth at most two. For this reason the functions $\FFpm{s}{m}{k}$ are referred to as depth-two modular invariant functions. 
 
Since the functions $\FFm{s}{m}{k}$ are odd under the involution $\tau \rightarrow - \bar \tau$ of the upper half-plane, they must be cuspidal, i.e., have a vanishing zeroth Fourier mode.
In the present paper we want to focus on reconstructing the non-perturbative $(q\bar{q})^n$ terms in the $0$-Fourier mode sector, but since for $\FFm{s}{m}{k}$ the $0$-mode sector identically vanishes, for the rest of the paper we will henceforth focus our attention purely on the $\FFp{s}{m}{k}$ and give some comments about the cuspidal $\FFm{s}{m}{k}$ in the conclusions.

The even modular-invariant functions $\FFp{s}{m}{k}$ are characterised by inhomogeneous Laplace eigenvalue 
equations that are very similar to the equations obeyed by lattice sums in the diagonalised basis (\ref{eq:diff_Cwmp}) namely
\begin{equation}
\label{int:Fmk}
\big(\Delta - s(s{-}1) \big) \FFp{s}{m}{k} = {\rm E}_m {\rm E}_k \, , \ \ \ \ \ \
s \in \left\lbrace k{-}m{+}2,k{-}m{+}4,\ldots,k{+}m{-}4,k{+}m{-}2 \right\rbrace \, ,
\end{equation}
where $2\leq m \leq k$, $\EE_m$, $\EE_k$ are non-holomorphic Eisenstein series, and $\Delta=4 (\Im \tau)^2 \partial_\tau \partial_{\bar \tau}$ is the $\SLtwoZ$ invariant Laplacian, as defined before. This differential equation fixes the asymptotics of $\FFp{s}{m}{k}$ at
the cusp $\tau \rightarrow i\infty$ up to two integration constants. The latter can be fixed from the
Poincar\'e-series representations whose seed functions enjoy shift symmetry 
under $\tau \rightarrow \tau{+}1$ or via alternative methods~\cite{Green:2008bf,DHoker:2017zhq}, which also fixes the boundary conditions for the differential equation. We note that the differential equation is invariant under the interchange of $s$ and $1{-}s$ and we always take $s$ to be greater than $1-s$. Furthermore, the equation is also invariant under the interchange of $m$ and $k$ and we label the function $\FFp{s}{m}{k}$ with $m\leq k$.

\subsection{Poincar\'e series approach}

A Poincar\'e series is a representation of a modular invariant function in terms of a sum over $\SLtwoZ$ images of another (modular non-invariant) function that we call its seed. Probably the simplest example of a Poincaré series comes from looking at the non-holomorphic Eisenstein series (\ref{eq:FEk}). These are well-known to be expressible as (see for example \cite{Fleig:2015vky})
\begin{equation}
    \label{eq:PSEk}
  \EE_w(\tau) = \frac{2\zeta(2w)}{\pi^w}\PS \Im(\gamma\cdot\tau)^w,
\end{equation}
where 
$B(\mathbb{Z}) = \big\{\big(\begin{smallmatrix}
  \pm 1 & n\\
  0 & \pm 1
\end{smallmatrix}\big) : n\in\mathbb{Z}\big\}$ 
is the Borel subgroup of translations. Observe that the seed function is just the monomial $(\Im\tau)^w$, so the procedure has reduced the depth by one unit. The Poincar\'e series~\eqref{eq:PSEk} converges for $\Re(w)>1$.

Similarly, we want to represent the modular invariant function $\FFp{s}{m}{k}$ as a sum over images of the group $\SLtwoZ$ of some seed function $f^{+(s)}_{m,k}$. It turns out that it is once again more convenient to choose the seed function to be periodic in the real direction and quotient out by the Borel subgroup in the sum. As a result we can write
\begin{equation}
\label{eq:Psum}
  \FFp{s}{m}{k}(\tau) =\PS f^{+(s)}_{m,k}(\gamma\cdot\tau),
\end{equation}
where $f^{+(s)}_{m,k}(\tau+1) = f^{+(s)}_{m,k}(\tau)$. 
The advantage for choosing the seed to be periodic comes from observing that it can then be Fourier decomposed as 
\begin{equation}
    \label{eq_f_fourier}
    f^{+(s)}_{m,k}(\tau) =  \sum_{n\in \mathbb{Z}}c_n(y) e^{2\pi i n \Re\tau},
\end{equation}
for real coefficient functions that satisfy $c_n(y) = c_{-n}(y)$, since we want a real-analytic function that is even under the involution $\tau\to-\bar{\tau}$.

Upon substitution of (\ref{eq_f_fourier}) into (\ref{int:Fmk}) we can ``fold''  $\EE_k$, i.e. use its known Poincaré sum representation \eqref{eq:PSEk}, to arrive at a simpler equation for the seed function
\begin{equation}
\label{eq_f_laplace}
    (\Delta - s(s-1))f^{+(s)}_{m,k} = (-1)^{k-1}\frac{\BB_{2k}}{(2k)!}(4y)^k \EE_m\,.
\end{equation}

From the known Fourier series of $\EE_m$ we can find an expression for the Fourier coefficients $c_n(y)$ that were defined in (\ref{eq_f_fourier}). These are Laurent polynomials in $y$ and matching corresponding powers on both sides of equation (\ref{eq_f_laplace}) gives an expression for $c_n(y)$. 
In \cite{Dorigoni:2021jfr,Dorigoni:2021ngn}, it was shown that the solution to this Laplace equation is given by

\begin{equation}
\begin{split}
\label{eq_c0_cn}
    c_0(y) = (-1)^{k+m}\frac{\BB_{2k}\BB_{2m}(4y)^{k+m}}{(2k)!(2m)!(\mu_{k+m}-\mu_s)} - (-1)^{k}\frac{4\BB_{2k}(2m-3)!\zeta_{2m-1}(4y)^{k+1-m}}{(2k)!(m-2)!(m-1)!(\mu_{k-m+1} - \mu_s)}\,, \\
    c_n(y) = (-1)^k\frac{2\BB_{2k}}{(2k)!\Gamma(m)}\sigma_{1-2m}(|n|)|n|^{m-k-1}\sum_{\ell=k-m+1}^{k-1}g^+_{m,k,\ell,s}(4|n|y)^\ell  e^{-2|n|y}\quad n\neq 0,
\end{split}
\end{equation}
where $\mu_s = s(s-1)$ and $g^+_{m,k,\ell,s}$ are the rational coefficients
\begin{equation}\label{eq:fkm}
    g^+_{m,k,\ell,s} = \frac{\Gamma(\ell)}{\Gamma(\ell+s)}\sum_{i=\ell}^{k-1}\frac{(\ell+1-s)_{i-\ell}\Gamma(s+i)\Gamma(m+k-i-1)}{\Gamma(k-i)\Gamma(i+1)\Gamma(m-k+i+1)},
\end{equation}
with $(a)_n = \frac{\Gamma(a+n)}{\Gamma(a)}$ the (ascending) Pochhammer symbol.

Equation~\eqref{eq_c0_cn} can be rewritten in a more suggestive way if we introduce one of the many flavours of iterated integrals at depth one:
\begin{align}
\label{eq:E0depth1}
\mathcal{E}_0(k,0^p;\tau) &= \frac{(2\pi i)^{p+1-k}}{p!}  \int_\tau^{i \infty}  (\tau{-}\tau_1)^p \GG_{k}^0(\tau_1)\dd\tau_1\,,
\end{align}
with even $k>2$ and the notation $0^p$ is a short-hand of $p$ successive zeros. Higher-depth versions, where the iterated integral structure becomes more evident, can be found in~\cite{Broedel:2015hia,Broedel:2018izr}. 

The symbol $\GG_{k}^0$ appearing in the integrand denotes the cuspidal part of the standard holomorphic Eisenstein series ${\rm G}_k(\tau) $:
\begin{align}
{\rm G}_k(\tau) &=\label{rev.12} \sum_{p \in \Lambda'} \frac{1}{p^k} 
= 2\zeta_{k} + \frac{2 (2\pi i)^k}{(k{-}1)!} \sum_{n>0} \sigma_{k-1}(n) q^n
\, , \ \ \ \ \ \ k \in\{ 4,6,8,...\} \, ,\\
{\rm G}_k^0(\tau) &\notag={\rm G}_k(\tau) - 2 \zeta_k
\end{align}
and it is convenient to define ${\rm G}_0^0=-1$.

The integral in (\ref{eq:E0depth1}) converges for $p \geq 0$ 
and from the $q$-expansion of~\eqref{rev.12} one can easily obtain~\cite{Broedel:2015hia, Dorigoni:2020oon}
\begin{align}
\mathcal{E}_0(k,0^p;\tau) &= -\frac{2}{(k{-}1)!}\sum_{m,n=1}^{\infty} \frac{m^{k-1}}{(mn)^{p+1}} q^{mn} 
=-\frac{2}{(k{-}1)!}\sum_{m=1}^{\infty} m^{k-p-2}\sigma_{1-k}(m) q^m  \label{eq:E0sigma} \\
&  =-\frac{2}{(k{-}1)!}\sum_{m=1}^{\infty} m^{-p-1}\sigma_{k-1}(m) q^m\,,
\notag
\end{align}
which can also be considered formally for arbitrary $k,p\in\mathbb{C}$.

Going back to the Fourier modes \eqref{eq_c0_cn} for the seed function $f^{+(s)}_{m,k}(\tau) $, we see that the general seed for all depth-two modular invariant functions can be written as
\begin{equation}
\label{eq_f_iterated}
    f^{+(s)}_{m,k}(\tau) = c_0(y) - (-1)^k\frac{2\BB_{2k}\Gamma(2m)}{(2k)!\Gamma(m)}\sum_{\ell=k-m+1}^{k-1}g^+_{m,k,\ell,s}(4y)^\ell\Re[\mathcal{E}_0(2m, 0^{k+m-\ell-1})].
\end{equation}

Noticeably, the use of Poincaré series has reduced the depth of the objects under consideration by one unit, thus making the problem more tractable. Furthermore, when $k>m$, the Poincar\'e seed just obtained gives rise to a convergent Poincar\'e sum.

Once the Poincar\'e seeds for the $\FFp{s}{m}{k}$ are known, we are also able to derive similar expressions for all two-loop MGFs, for example~\cite{Dorigoni:2019yoq},
\begin{equation}
    \label{eq_C211}
    C_{2,1,1}(\tau) = \PS \bigg[\frac{2y^4}{14175}+\frac{y\zeta(3)}{90}+\frac{y}{90}\sum_{m=0}^\infty \sigma_{-3}(m)(q^m+\bar{q}^m)\bigg]_{\gamma},
\end{equation}
where $[...]_{\gamma}$ implies an action of $\gamma$ on everything in the bracket. Once again we note that a perk of using such a Poincar\'e series representation is that the depth of the MGF was reduced by one, since the sum in the brackets is related to the depth-1 object $\EE_2(\tau)$ through its Fourier series \eqref{eq:FEk}. 

\section{Resurgent analysis for Poincar\'e series}
\label{sec:3}

The task at hand is now to start from the Poincar\'e-series representation \eqref{eq:Psum} in terms of seed functions and extract the asymptotic expansion at the cusp of the modular-invariant function $ \FFp{s}{m}{k}(\tau)$. 

We can consider again the Eisenstein series as a warm-up exercise, and very standard results~\cite{Iwaniec:2002,Fleig:2015vky} tell us how to obtain the asymptotic expansion at the cusp~\eqref{eq:FEk} from its Poincar\'e sum representation~\eqref{eq:PSEk}.
For more general Poincar\'e series the analysis is more involved, but in principle it is possible to rewrite each Fourier coefficient of a modular invariant function in terms of some convoluted integral transform of the Fourier coefficients of its seed function as well as involving complicated Kloosterman sums. We review this general procedure in appendix~\ref{app:Poincare}.

In the present case we can see that the non-zero Fourier mode of the general seed \eqref{eq_c0_cn} is of the form 
\begin{equation}
c_n(y) = \sum_{\ell=k-m+1}^{k-1}\Big[ (-1)^k\frac{\BB_{2k}}{(2k)!\Gamma(m)} g^+_{m,k,\ell,s}\Big] \sigma_{2m-1}(|n|) |n|^{\ell-k-m} (4y)^{\ell} e^{-2|n|y}\,,\label{eq:seednclean}
\end{equation}
hence a finite and rational linear combination of seeds of the type
\begin{equation}
\label{eq:genseed}
\sigma_a(|n|) |n|^b y^r e^{-2|n|y}\,.
\end{equation}

Seeds of precisely this form were studied in~\cite{Ahlen:2018wng,Dorigoni:2019yoq}, where it was shown how to use the procedure outlined in Appendix \ref{app:Poincare} to compute the Laurent polynomial of the associated Poincar\'e sum. To summarise the result, we consider the Poincar\'e sum
\begin{equation}
\summ(\tau) = \sum_{\ell \in\mathbb{Z}} a_\ell (y) e^{2\pi i \ell \Re \tau} = \PS \seeed(\gamma \tau)\,,
\end{equation}
with seed function given by terms of the form~\eqref{eq:genseed}
\begin{align}
\seeed(\tau) &\label{eq:defseed}= \sum_{n\in\mathbb{Z}\setminus\{0\}} c_n(y) e^{2\pi i n \Re \tau}\,,\\
c_n (y) &\notag= \sigma_{a}(\vert n \vert) \vert n \vert ^b y^r e^{-2 \vert n \vert y}\,.
\end{align}
Then the Laurent polynomial part of the asymptotic expansion at the cusp $y\to\infty$, for the zero-mode coefficient $a_0(y)$ is given by
\begin{align}
a_0(y)\sim I(a,b,r;y) &= \frac{2^{3-2r} y^{1+b-r}}{\Gamma(r)\pi^{2b-2r}}\Bigg[ \frac{y}{\pi^2} \frac{\Gamma(b{+}1)\Gamma(2r{-}b{-}2)}{\Gamma(r{-}b{-}1)} \frac{\zeta(2r{-}a{-}2b{-}2)\zeta(1{-}a)}{\zeta(2r{-}a{-}2b{-}1)} \nn\\
&\hspace{2mm} + \left(\frac{y}{\pi^2}\right)^{a+1} \frac{\Gamma(a{+}b{+}1)\Gamma(2r{-}a{-}b{-}2)}{\Gamma(r{-}a{-}b{-}1)}\frac{\zeta(2r {-}a{-}2b{-}2)\zeta(a{+}1)}{\zeta(2r{-}a{-}2b{-}1)} \nn\\
&\hspace{2mm} +\left(\frac{\pi^2}{y}\right)^b 
\sum_{n \geq 0} \left(\frac{-\pi^2}{y}\right)^{n} \frac{\Gamma(2r{+}n{-}1)}{n! \cdot \Gamma(r{+}n)}  \label{eq_a0_asymp}\\
&\hspace{20mm}\times \frac{\zeta({-}b{-}n)\zeta({-}a{-}b{-}n)\zeta(2r{-}a{-}b{+}n{-}1)\zeta(2r{-}b{+}n{-}1)}{\zeta(2r{+}2n)\zeta(2r{-}a{-}2b{-}1)} \bigg]\,.\notag
\end{align}

Using Ramanujan's identity
\begin{equation}
\sum_{n=1}^\infty  \frac{\sigma_a(n)\sigma_b(n)}{n^s} = \frac{\zeta(s)\zeta(s{-}a)\zeta(s{-}b)\zeta(s{-}a{-}b)}{\zeta(2s{-}a{-}b)}\,,\label{eq:Rama}
\end{equation}
the last term can be rewritten in Dirichlet series form
\begin{align}
&\quad     \frac{\zeta(-b-n)\zeta(-a-b-n)\zeta(2r-a-b+n-1)\zeta(2r-b+n-1)}{\zeta(2r+2n)} \nn\\
&= 4\sin{\left(\frac{\pi(b{+}n)}{2}\right)}\sin{\left(\frac{\pi(a{+}b{+}n)}{2}\right)}\frac{\Gamma(1+b+n)\Gamma(1+a+b+n)}{(2\pi)^{a+2b+2n+2}}\sum_{m>0}\frac{\sigma_a(m)\sigma_{a+2b+2-2r}(m)}{m^{a+b+n+1}}\,.\label{eq:Rama2}
\end{align}

A few comments regarding the general expression \eqref{eq_a0_asymp} are in order.
\begin{itemize}
\item For generic $a,b,r$ this asymptotic series is a Gevrey-1, factorially divergent formal power series. Shortly we will use Borel resummation in order to reconstruct the non-perturbative properties of $a_0(y)$ at the cusp $y\to \infty$. As usual, ambiguities in prescribing a unique resummation procedure will allow us to obtain the exponentially suppressed contributions, $(q\bar{q})^n$, which are hidden in the purely perturbative asymptotic result \eqref{eq_a0_asymp}.
\item
For $a,b$ integers with $a$ odd (as for the case under consideration \eqref{eq:seednclean}), the series in \eqref{eq_a0_asymp} terminates after a finite number of terms.
This can be easily understood by noticing that for $n$ large enough either $\zeta(-b-n)$ or $\zeta(-a-b-n)$ will be a zeta value at a negative even integer, hence vanishing, while all other factors will be regular. For $a,b$ integers with $a$ odd we then have that the series in \eqref{eq_a0_asymp} does terminate for $n>n_{max} = {\rm{max}}(-b,-a-b)+1$. In particular for our choice of seeds \eqref{eq:seednclean}, we have $a=2m-1\geq 0 $ while $-b\in\{m+1,m+2,...,2m,2m+1\}$, hence for the case of interest \eqref{eq_a0_asymp} always truncates for $n> -b+1$.

\item The parameter $b$ serves the purpose of a regulator. When $b$ is arbitrary our expression \eqref{eq_a0_asymp} is a formal asymptotic power series for which we can make use of resurgent analysis to reconstruct the exponentially suppressed terms in the $0$-mode. At the end of the day, when we set $b$ to its physical values appearing in \eqref{eq:seednclean} the asymptotic power series will truncate to the expected finite Laurent polynomial, while the non-perturbative terms will survive. This is an instance of Cheshire cat resurgence.
\end{itemize}

In \cite{Dorigoni:2021jfr} it was indeed shown that if we use the general expression \eqref{eq_a0_asymp} specialised to the seed  $\seedp{s}{m}{k}$ from \eqref{eq_f_iterated} then we obtain a truncating Laurent polynomial for the modular invariant functions $ \FFp{s}{m}{k}(\tau)$:
\begin{align}
\FFp{s}{m}{k} &= {\rm{P}}_{m,k}^{(s)}(y)+ O(q,\bar{q})\,,
\notag
\end{align}
where the Laurent polynomial ${\rm{P}}_{m,k}^{(s)}(y)$ is given by
\begin{align}
{\rm{P}}_{m,k}^{(s)}(y) \! =& \frac{(-4)^{k{+}m} {\rm B}_{2m} {\rm B}_{2k} }{(k{+}m{-}s)(k{+}m{+}s{-}1) (2m)! (2k)!} y^{k+m}
\!-\!   \frac{2(-1)^{m} 4^{1{+}m{-}k}{\rm B}_{2m}  \Gamma(2k{-}1) \zeta_{2k{-}1}}{\Gamma(k) \Gamma(k) (m{-}k{+}s)(m{-}k{-}s{+}1)(2m)!} y^{1+m-k}
\nn\\
&\quad 
-   \frac{2(-1)^{k} 4^{1{+}k{-}m}{\rm B}_{2k}  \Gamma(2m{-}1) \zeta_{2m{-}1}}{\Gamma(m) \Gamma(m) (k{-}m{+}s)(k{-}m{-}s{+}1)(2k)!} y^{1+k-m}
\label{eq:FLP}\\
&\nn\quad +  \frac{4^{3-m-k}\Gamma(2m{-}1)\Gamma(2k{-}1)\zeta_{2m{-}1}\zeta_{2k{-}1}}{[\Gamma(m)\Gamma(k)]^2 (k{+}m{-}s{-}1)(k{+}m{+}s{-}2)} y^{2-k-m}
+ c_{m,k}^{(s)} \zeta_{k+m+s-1} y^{1-s}\,,
\end{align}
with the rational coefficient
\begin{align}
 c_{m,k}^{(s)} = \frac{4^{2-s} (-1)^{m+s+1} {\rm B}_{s+m-k} {\rm B}_{k+m-s} {\rm B}_{k+s-m}(2s)! }{(s{+}m{-}k)\Gamma(m)\Gamma(s) {\rm B}_{2s} (k{+}m{-}s)! (k{+}s{-}m)!} \sum_{\ell=k-m+1}^{\min(k-1,s)} (-1)^\ell g^+_{m,k,\ell,s} \frac{\Gamma(\ell{+}s{-}1)}{\Gamma(\ell) (s{-}\ell)!}\,,
\end{align}
expressed in terms of the rational numbers $g^+_{m,k,\ell,s}$ defined in~\eqref{eq:fkm}.

The last term in \eqref{eq:FLP} satisfies the homogeneous Laplace equation \eqref{int:Fmk} and its coefficient can also be rewritten \cite{Green:2008bf}\footnote{In this reference the coefficient of the homogeneous solution is determined by projecting the Laplace equation \eqref{int:Fmk} on ${\rm E}_{s}(\tau)$ and integrating over the fundamental domain.} as
\begin{equation}
 c_{m,k}^{(s)}  =- 4 \pi^{\frac{s-m-k-1}{2}}\frac{\Gamma\Big(\frac{m+k+s-1}{2}\Big) \zeta^*(s+1-m-k)\zeta^*(m+s-k)\zeta^*(k+s-m)}{
(2s-1)  \Gamma(m)\Gamma(k)\,\zeta^*(2s)},
\end{equation}
with $\zeta^*(s)=\zeta(s)\Gamma(\frac{s}{2})\pi^{-\frac{s}{2}}$.

\subsection{Resumming an evanescent tail}
\label{sec:3.1}

Since we are interested in exploiting the asymptotic nature of the general expression (\ref{eq_a0_asymp}), we can simply focus on its last term which, for generic $a,b,r$, does indeed produce the factorially divergent asymptotic tail
\begin{align}
    I_{asy}(a,b,r ; y) &\label{I_asy}= \frac{(4y)^{2+a+b-r}\pi^{2r-a-2b-2}}{2^{a+2b}\Gamma(r)\zeta(2r-a-2b-1)}\sum_{n>0}\sigma_a(n)\sigma_{a+2b+2-2r}(n)\\
   &\nn \!\!\!\!\!\!\!\!\!\! \!\!\!\!\! \sum_{m\geq 0}\frac{\Gamma(m+a+b+1)}{(4 ny)^{m+a+b+1} } \frac{\Gamma(2r+m-1)\Gamma(1+b+m)}{\Gamma(m+r)\Gamma(m+1)}\Big[(-1)^m\cos\Big(\frac{a\pi}{2}\Big) - \cos\Big(\frac{(a+2b)\pi}{2}\Big)\Big]\,,
\end{align}
after making use of Ramanujan's identity as discussed above.
We note that $I_{asy}(a,b,r;y)$ should be understood only as a formal power series in $y^{-1}$ with zero radius of convergence.

The next step is to perform a standard Borel resummation for \eqref{I_asy}. 
Rewriting the integral representation of the gamma function as
\begin{equation}
\frac{\Gamma(m+a+b+1)}{(4ny)^{m+a+b+1}} = \int_0^\infty e^{-4 n y t} t^{m+a+b} \dd t\,,
\end{equation}
we can then define the directional Borel resummation, see \cite{Dorigoni:2014hea} for a recent review, of the formal power series $  I_{asy}(a,b,r;y) $ as
\begin{align}
 & \mathcal{S}_\theta \Big[I_{asy} (a,b,r;y)\Big]=\\
 &\nn \frac{(4y)^{2+a+b-r}\pi^{2r-a-2b-2}}{2^{a+2b}\Gamma(r)\zeta(2r-a-2b-1)}\frac{\Gamma(2r-1)\Gamma(1+b)}{\Gamma(r)} \sum_{n>0}\sigma_a(n)\sigma_{a+2b+2-2r}(n) \int_0^{e^{i\theta}\infty} e^{-4 nyt}B(t) \dd t\,,
\end{align}
where the Borel transform in the case at hand is given by 
\begin{align}
B(t) &\label{Borel_transf_Iasy} = \sum_{n\geq 0}t^{a+b+n} \frac{(2r-1)_n(1+b)_n}{(r)_n n!}\Big[(-1)^n\cos{\Big(\frac{a\pi}{2}\Big)} - \cos{\Big(\frac{(a+2b)\pi}{2}}\Big)\Big]\\
&\nn=  t^{a+b}\Big[{}_2F_1(2r-1,1+b;r\vert -t)\cos{\Big(\frac{a\pi}{2}\Big)} - {}_2F_1(2r-1,1+b;r\vert t)\cos{\Big(\frac{(a+2b)\pi}{2}}\Big)\Big]\,,
\end{align}
with ${}_2F_1(a,b;c\vert z)$ denoting a standard hypergeometric function.

We see that for $\theta \in (0,\pi/2)$, the directional Borel resummation $ \mathcal{S}_\theta \Big[I_{asy} (a,b,r;y)\Big]$ does indeed define an analytic function in the complex wedge $-\pi/2-\theta< \mbox{arg} \,y <\pi/2 -\theta$ whose asymptotic expansion near $y\to\infty$ is precisely given by \eqref{I_asy}.
Furthermore, if we take two different directions $\theta_1,\theta_2\in (0,\pi/2)$, with $\theta_1<\theta_2$, it is simple to see that $ \mathcal{S}_{\theta_1} \Big[I_{asy} (a,b,r;y)\Big]$ and $\mathcal{S}_{\theta_2} \Big[I_{asy} (a,b,r;y)\Big]$ are analytic continuations of one another since the integrand is regular in the complex wedge $\theta_1\leq \mbox{arg}\,t\leq \theta_2$.

A similar story can be repeated for $\theta \in (-\pi/2,0)$, however if we define the lateral Borel resummation as
\begin{equation}
\mathcal{S}_\pm \Big[I_{asy} (a,b,r;y)\Big] = \lim_{\theta\to 0^\pm} \mathcal{S}_{ \theta}\Big[I_{asy} (a,b,r;y)\Big] \,,
\end{equation}
we see that the two analytic continuations $\mathcal{S}_\pm \Big[I_{asy} (a,b,r;y)\Big] $ differ on the common domain of analyticity, since the integrand, and in particular ${}_2F_1(2r-1,1+b;r \vert t)$, has a branch-cut singularity precisely along the direction $\mbox{arg}\,t=0$, hence called a Stokes direction.
Thus we have obtained two analytic continuations of the same formal power series $\eqref{I_asy}$ which differ precisely on the direction of interest, namely $y>0$. This is generically a signal that we have to include non-perturbative, exponentially suppressed corrections~\cite{Dorigoni:2014hea}.

From the properties of the hypergeometric series we can easily compute its discontinuity across the branch cut $t\in[1,\infty)$:
\begin{align}
  &\quad \label{eq:disc2F1} \mbox{Disc}_0\Big[{}_2F_1(a,b;c \vert t)\Big]= \lim_{\epsilon\to 0}\Big[{}_2F_1(a,b;c \vert t+i\epsilon) - {}_2F_1(a,b;c \vert t-i\epsilon)\Big]\nn\\
  &= \frac{2\pi i\Gamma(c)}{\Gamma(a)\Gamma(b)\Gamma(c-a-b+1)}(t-1)^{c-a-b}{}_2F_1(c-a,c-b;c-a-b+1\vert 1-t)\,,
\end{align}
valid for $t>1$.
We can then compute the difference between the two lateral resummations, related to what is called the Stokes automorphism, and find
\begin{align}
\label{eq_Stokes}
 & \quad (\mathcal{S}_+ - \mathcal{S}_-)\Big[I_{asy}(a,b,r;y)\Big] \\
 &\nn= \frac{(4y)^{2+a+b-r}\pi^{2r-a-2b-2}}{2^{a+2b}\Gamma(r)\zeta(2r-a-2b-1)}\frac{\Gamma(2r-1)\Gamma(1+b)}{\Gamma(r)} \sum_{n>0}\sigma_a(n)\sigma_{a+2b+2-2r}(n) \!\int_0^{\infty}\!\!\! e^{-4 nyt}\mbox{Disc}_0B(t) \dd t\\
&\nn=- \frac{(4y)^{2+a+b-r}\pi^{2r-a-2b-2}}{2^{a+2b}\Gamma(r)\zeta(2r-a-2b-1)}\sum_{n>0}\sigma_a(n)\sigma_{a+2b+2-2r}(n) 2\pi i \cos{\Big(\frac{(a+2b)\pi}{2}\Big)} e^{-4 ny}\\
&\nn\phantom{=}\times\int_0^\infty e^{-4 nyt}\frac{(t+1)^{a+b}  t^{-r-b}}{\Gamma(1-r-b)}{}_2F_1(1-r,r-b-1;1-r-b\vert -t)\dd t\,,
\end{align}
where in the last step we substituted the discontinuity \eqref{eq:disc2F1} and shifted the integration variable $t\to t+1$.

Notice that this discontinuity in resummation is purely non-perturbative in nature due to the presence of the exponentially suppressed term $(q\bar{q})^n = e^{-4ny}$.
The present discussion is very similar to \cite{Arutyunov:2016etw,Dorigoni:2019yoq,Dorigoni:2020oon}: the starting asymptotic
series \eqref{eq_a0_asymp} cannot be easily Borel resummed as it is, however by realising that the factorially growing coefficients are “dressed” by  a suitable Dirichlet series we obtain \eqref{I_asy}, amenable to standard Borel resummation. The infinitely many exponentially suppressed corrections $(q\bar{q})^n = e^{-4ny}$ can be seen as arising from the unfolding of the Dirichlet series \eqref{eq:Rama2} combined with the shift $y\to 4 n y$.

The median resummation of the asymptotic formal power series $I_{asy}(a,b,r;y)$ is then defined by
\begin{equation}\label{eq:med}
\mathcal{S}_{med}\Big[I_{asy}(a,b,r;y)\Big] = \mathcal{S}_{\pm} \Big[I_{asy}(a,b,r;y)\Big] \mp i \,\mbox{Im}[\sigma(a,b)]\, \mbox{NP}(a,b,r;y)\,,
\end{equation}
which is independent of our choice of sign, i.e. of direction of resummation, 
having defined the imaginary part of the transseries parameter
\begin{equation}
\label{eq:NP1}
\mbox{Im}\,[\sigma(a,b)]= \cos\Big(\frac{(a+2b)\pi}{2}\Big)\,,
\end{equation} and the non-perturbative part $\mbox{NP}(a,b,r;y)$ is given
\begin{align}
\mbox{NP}(a,b,r;y)& = - \frac{(4y)^{2+a+b-r}\pi^{2r-a-2b-1}}{2^{a+2b}\Gamma(r)\zeta(2r-a-2b-1)}\sum_{n>0}\sigma_a(n)\sigma_{a+2b+2-2r}(n)  e^{-4 ny}\\
&\nn\phantom{=}\times\int_0^\infty e^{-4 nyt} (t+1)^{a+b}  t^{-r-b} {}_2\tilde{F}_1(1-r,r-b-1;1-r-b\vert -t)\dd t\,,
\end{align}
where ${}_2\tilde{F}_1(a,b;c\vert z) = {}_2F_1(a,b;c\vert z)/\Gamma(c)$ denotes the regularised hypergeometric function.

We notice that the discontinuity \eqref{eq_Stokes} and in particular the Stokes constant $\cos [(a+2b)\pi/2 ]$, only fixes the imaginary part of the transseries parameter $ \sigma(a,b)$, i.e. the overall piece-wise constant (jumping only at Stokes directions) in front of the non-perturbative
terms.
Following \cite{Dorigoni:2019yoq,Dorigoni:2020oon}, we will make the assumption that the complete transseries
parameter does in fact depend analytically on $(a+2b)$, and the “minimal analytic completion” with non-trivial real part is simply
\begin{align}\label{eq:tsexp}
\sigma_\pm(a,b) &= \exp\Big(\pm i\pi\frac{a+2b-1}{2}\Big) = \sin\Big(\frac{(a+2b)\pi}{2}\Big) \mp i \cos\Big(\frac{(a+2b)\pi}{2}\Big) 
\end{align}
where once more we stress that the sign $\pm$ is correlated with the choice of resummation as in \eqref{eq:med}.

Usually when we look for transseries solutions to say non-linear ODEs, the imaginary part of the transseries parameter is fixed by the Stokes discontinuity, while its real part is determined via some initial condition. At the present time we do not have such an ODE construction for our problem and we are in a certain sense trying to bootstrap the full transseries entirely out of the perturbative data generated by the Poincar\'e sum of our seed \eqref{eq:defseed} for $a,b\in\mathbb{C}$ generic, without having at our disposal any ODE or functional equation to guide us. 

One of the key features of what is generally called ''Cheshire cat resurgence'' \cite{Dunne:2016jsr,Kozcaz:2016wvy,Dorigoni:2017smz,Dorigoni:2019kux} is precisely that the Stokes constant crucially vanishes for special values of the deformation parameter ($a+2b$ in the present case or a supersymmetry breaking deformation in the aformentioned references) while non-perturbative corrections are expected to be present for all values of the deformation. This implies that the transseries parameter must have a non-vanishing real part as well. Our hypothesis  \eqref{eq:tsexp} provides the minimal analytic completion to achieve this, and, as we will see later on, will produce the correct non-perturbative terms.

We then conclude that the non-perturbative resummation of \eqref{I_asy} is given by
\begin{equation}\label{eq:med2}
\mathcal{S}_{med}\Big[I_{asy}(a,b,r;y)\Big] = \mathcal{S}_{\pm} \Big[I_{asy}(a,b,r;y)\Big]+\sigma_\pm(a,b)  \mbox{NP}(a,b,r;y)\,.
\end{equation}
Thanks to the discontinuity equation \eqref{eq_Stokes}, we can easily see that the this is a well-defined analytic function providing a non-perturbative and unambiguous resummation for the formal asymptotic power series \eqref{I_asy} which is also real (as one would have expected) for $y>0$ with $a,b\in\mathbb{R}$ and continuous as $\mbox{arg} \,y\to0$.

\begin{figure}
    \centering
    \includegraphics[scale=0.5]{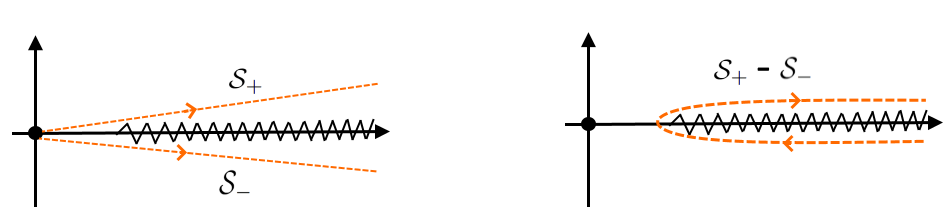}
    \caption{\label{fig:my_label}\textit{On the left diagram we show the two different lateral Borel resummation. On the right diagram, the difference between the two lateral Borel resummation is represented as an Hankel integral contour, used to evaluate the Stokes automorphism.}}
\end{figure}

\subsection{Non-perturbative completion}

We can now specialise the results of the previous section for the generic seed \eqref{eq:defseed} to the case of interest for the seed functions $f^{+(s)}_{m,k}$ relevant for all two-loop MGFs.
We can rewrite~\eqref{eq_f_iterated} as 
\begin{equation}
\label{eq_f_seed_combo}
    f^{+(s)}_{m,k}(y) = c_0(y) + (-1)^k\frac{2\BB_{2k}}{(2k)!\Gamma(m)}\sum_{n\neq 0} \sum_{\ell=k-m+1}^{k-1}g^+_{m,k,\ell,s}(4y)^\ell|n|^{\ell-k-m}\sigma_{2m-1}(|n|)e^{-2|n|y}e^{2\pi in\tau_1}\,,
\end{equation}
such that, manifestly, each seed is a finite combination of building blocks \eqref{eq:defseed} with $a=2m-1$, $b=\ell-k-m$, $r=\ell$ just analysed.

We start by decomposing $\FFp{s}{m}{k}$ in Fourier modes for $\tau_1 =\mbox{Re}\,\tau$
\begin{equation}\label{eq:FourierExp}
\FFp{s}{m}{k}(\tau) = \sum_{n\in\mathbb{Z}} a_n(m,k; s; y) e^{2\pi i n\tau_1}\,,
\end{equation}
and focus on the asymptotic expansion for $y\to\infty$ of the zero-mode $a_0(m,k;s;y)$\,.

From the seed mode expansion \eqref{eq_f_seed_combo} we can use the results of Appendix \ref{app:Poincare} to arrive at the Laurent polynomial part
\begin{align}
  &\nn  a_0(m,k;s;y) \sim\\
  &\nn (-1)^{k+m}\frac{\BB_{2k}\BB_{2m}4^{k+m}I_0(k+m;y)}{(2k)!(2m)!(\mu_{k+m}-\mu_s)} - (-1)^{k}\frac{4\BB_{2k}(2m-3)!\zeta_{2m-1}I_0(k+1-m;y)}{(2k)!(m-2)!(m-1)!(\mu_{k-m+1} - \mu_s)}\\
  &\label{eq:a0formal}+  (-1)^k\frac{2\BB_{2k}}{(2k)!\Gamma(m)} \sum_{\ell=k-m+1}^{k-1}g^+_{m,k,\ell,s} 4^{\ell} I(2m-1,\ell-k-m,\ell;y)\,,
  \end{align}
  where $I_0(r;y)$, defined in \eqref{eq:I0}, comes from the Poincar\'e sum of the seed function zero-mode $c_0(y)$ \eqref{eq_c0_cn}, while $I(a,b,r;y)$ \eqref{eq_a0_asymp} comes from the Poincar\'e sum of the non-zero modes.
 
As explained in \cite{Dorigoni:2019yoq,Dorigoni:2021jfr,Dorigoni:2021ngn}, there are a few instances where the above expression has to be regulated. For example, it is fairly easy to see from \eqref{eq:I0} that whenever $k=m$ the second contribution, naively proportional to $I_0(1;y)$, is divergent.
The correct way to proceed is to regulate this expression by  shifting $k\to k+\epsilon$ where the expression becomes regular for all $m,k\geq 2$. 
To render the whole expression regular, it is actually enough to consider the regulator $I_0(k+1-m;y)\to I_0(k+\epsilon+1-m;y)$ and $ I(2m-1,\ell-k-m,\ell;y)\to I(2m-1,\ell-k-\epsilon-m,\ell;y)$.

The $I(a,b,r;y)$ contribution, coming from the Poincar\'e sum of all the non-zero Fourier modes of the seed function, gets specialised to the particular values $I(2m-1,\ell-k-m,\ell;y)$, hence the regulator $k\to k+\epsilon$ amounts to considering an analytic continuation in the $b$ parameter, thus introducing an asymptotic tail of factorially growing terms just as discussed in the previous section. 

As we send $\epsilon \to 0$ this tail will disappear and the precise combination of $I_0(r;y)$ and $I(a,b,r;y)$ contributions will give rise to the finite Laurent polynomial \eqref{eq:FLP}. However as argued above, the non-perturbative terms needed to provide an unambiguous resummation of the formal power series for $\epsilon \neq 0$ will survive in this limit, thus giving us the full, non-perturbative zero-mode contribution $a_0(m,k;s;y)$ to $\FFp{s}{m}{k} (\tau)$.

In more detail, we can consider the non-perturbative resummation \eqref{eq:med2} for the formal power series $I_{asy}(a,b,r)$ and specialise it to the current case \eqref{eq:a0formal} arriving at:
\begin{align}
  &\label{eq:0modefinal}  a_0(m,k;s;y) = {\rm{P}}_{m,k}^{(s)}(y)+\\
  &\nn  \lim_{\epsilon\to0}\Big\lbrace(-1)^k\frac{2\BB_{2k}}{(2k)!\Gamma(m)} \!\! \sum_{\ell=k-m+1}^{k-1}\!\! g^+_{m,k,\ell,s} 4^l \mathcal{S}_\pm \big[ I_{asy}(2m-1,\ell-k-\epsilon-m,\ell;y)\big]\!+\! {\rm{NP}}^\epsilon_\pm(m,k;s;y)  \Big\rbrace 
  \end{align}
where we collected in $ {\rm{P}}_{m,k}^{(s)}(y)$ all the regular and finitely many perturbative terms arising from the limit for $\epsilon\to0$ of \eqref{eq:a0formal} and which reproduce the Laurent polynomial \eqref{eq:FLP}.
When $\epsilon$ is sent to zero we know, from our discussion above, that the resummation of the asymptotic tail $I_{asy}$ vanishes identically, i.e. there is no asymptotic tail when $\epsilon=0$.
Finally the non-perturbative terms, which will survive in the $\epsilon\to0$ limit, are given by
\begin{align}
   \label{eq_F_Stokes}
&{\rm{NP}}^\epsilon_\pm(m,k;s;y) = \\
&\nn \frac{2\times 4^{2+m-k}y^{m+1-k}}{\Gamma(m)}\sum_{n>0}\sigma_{1-2k}(n)\sigma_{2m-1}(n)e^{-4ny} \sum_{\ell=k-m+1}^{k-1}\frac{g^+_{m,k,\ell,s}e^{\pm i \pi (\ell-k-\epsilon)}}{\Gamma(\ell)}\\
&\nn \times \int_0^\infty e^{-4nyt} (t+1)^{\ell+m-k-1}\,t^{k+m+\epsilon-2\ell}\,{}_2\tilde{F}_1(1-\ell,k+m-1;k+m+1+\epsilon-2\ell\vert -t) \dd t\,,
\end{align}
where ${}_2\tilde{F}_1(a,b;c\vert z)$ is the regularised hypergeometric function, see~\eqref{eq:NP1}.
The suffix $\pm$ is a reminder that we have already specialised the transseries parameter $\sigma_\pm(a,b)$ from \eqref{eq:tsexp} to the present case $a=2m-1, b= \ell-k-m-\epsilon$:
\begin{equation}\label{eq:TSspec}
\sigma_\pm(2m-1, \ell-k-m-\epsilon) = e^{\pm i \pi (\ell-k-\epsilon-1)}\stackrel{\epsilon\to0}{\longrightarrow} (-1)^{\ell+k+1}\,.
\end{equation}

The main result of our paper is given by equation (\ref{eq_F_Stokes}) which contains all the exponentially suppressed $(q\bar{q})^n = e^{-4ny}$ terms in the zero-mode sector of all depth-two modular invariant functions $\FFp{s}{m}{k}$. 

The key role of the parameter $\epsilon$ is to regulate the Borel transform integrand in equation \eqref{eq_F_Stokes}.
To make things clearer, let us analyse the various terms appearing in the integrand and see what the regulator $\epsilon$ does.

Firstly, we see from \eqref{eq:TSspec} that the transseries parameter is perfectly regular in this limit and it reduces to $(-1)^{\ell+k+1}$.
Secondly, for the range of parameters considered here, the term $(t+1)^{\ell+m-k-1}$ is simply a polynomial in $t$ of degree at most $m-2$. Similarly, the regularised hypergeometric function ${}_2\tilde{F}_1(1-\ell,k+m-1; k+m+1+\epsilon-2\ell\vert -t)$ is also a polynomial in $t$ of degree $\ell-1$, since its first entry is a non-positive integer while the third entry is generic due to the presence of the regulator $\epsilon$.

Hence we arrive at the conclusion that the integrand can be written as a polynomial in $t$ multiplied by $t^{k+m+\epsilon-2\ell}$, a non-integer power of $t$, and the usual exponential damping factor. For generic $\epsilon$, each monomial in $t$ can be easily integrated to produce a gamma function multiplied by a power of $(4ny)$, i.e.
\begin{equation}
\int_0^{\infty} e^{-4 n y t} t^{k+m+\epsilon-2\ell} t^n \dd t = \frac{\Gamma(k+m+n+1+\epsilon-2\ell) }{(4ny)^{k+m+n+1+\epsilon-2\ell}}\,.\label{eq:gamma}
\end{equation} 

We need to distinguish two cases now:
\begin{itemize}
    \item When $2\ell\leq k+m$, the regulating factor $t^{k+m+\epsilon-2\ell}$ is a positive power of $t$ and the integral is regular in the limit $\epsilon\to0$, hence we can directly compute:
    \begin{equation}
    \int_0^\infty e^{-4nyt} (t+1)^{\ell+m-k-1}\,t^{k+m-2\ell}\,{}_2\tilde{F}_1(1-\ell,k+m-1;k+m+1-2\ell\vert -t) \dd t\,,
    \end{equation}
    which is a polynomial of degree $2m-1$ in $(4 n y)^{-1}$.
    \item  When $2\ell\geq k+m+1$, the regulating factor $t^{k+m+\epsilon-2\ell}$ is a negative power of $t$ making the integral ill-defined in the strict $\epsilon=0$ limit. However, in this case, the regularised hypergeometric series ${}_2\tilde{F}_1(1-\ell,k+m-1; k+m+1+\epsilon-2\ell\vert -t)$ has a negative third entry. If we write the hypergeometric in terms of Gauss' series
    \begin{equation*}
    {}_2\tilde{F}_1(1-\ell,k+m-1; k+m+1+\epsilon-2\ell\vert-t) = \sum_{n=0}^{\infty} \frac{(1-\ell)_n (k+m-1)_n}{\Gamma(k+m+1+\epsilon+n-2\ell)} \frac{(-t)^n}{n!}\,,
    \end{equation*}
    it is now manifest that for $0\leq n \leq 2\ell - k - m$ the coefficient of $t^n$ vanishes in the $\epsilon\to0$ limit, being proportional to $\Gamma(k+m+1+\epsilon+n-2\ell)^{-1}$.
This vanishing behaviour exactly cancels the divergence that would originate from integrating, as in \eqref{eq:gamma}, any negative power of $t$ generated from the factor $t^{k+m+\epsilon-2\ell}$ term. We find that again there is a well-defined limit $\epsilon\to 0$, that has to be taken \textit{after} having performed the $t$-integral:
    \begin{equation*}
   \lim_{\epsilon\to0} \Big[\int_0^\infty e^{-4nyt} (t+1)^{\ell+m-k-1}\,t^{k+m+\epsilon-2\ell}\,{}_2\tilde{F}_1(1-\ell,k+m-1;k+m+1+\epsilon-2\ell\vert -t)\dd t \Big] \,,
    \end{equation*}
 which is again a polynomial of degree $2m-1$ in $(4 n y)^{-1}$.
\end{itemize}
This concludes a proof that equation (\ref{eq_F_Stokes}) is regular as $\epsilon\to 0$. This limit is interpreted as describing the non-perturbative corrections to the Fourier zero mode of $\FFp{s}{m}{k}$ through resurgent analysis and we have recovered the exact behaviour of the zero-mode at the cusp.

\subsection{Some Examples}
\label{sec:3.2}

We list some of the results for the zero-mode $a_0(m,k;s;y)$ that follow from the previously derived calculations for a few small values of $m,k,s$.

 {\allowdisplaybreaks 
In the $(2,k)$ sector, where there is a single eigenvalue $s=k$, we have:
\begin{align}\label{eq_examples2k}
    a_0(2,2;2;y)&= \frac{y^4}{20250}-\frac{y\zeta_3}{45} - \frac{5\zeta_5}{12y} + \frac{\zeta_3^2}{4y^2} + \sum_{n=1}^\infty\frac{e^{-4ny}\sigma_{-3}(n)^2}{2y^2}\,,\\\label{eq_examples2k_2}
          a_0(2,3;3;y)&= \frac{y^5}{297675}-\frac{y^2\zeta_3}{1890} - \frac{\zeta_5}{360} -\frac{7\zeta_7}{64y^2} + \frac{\zeta_3\zeta_5}{8y^3} + \sum_{n=1}^\infty e^{-4ny}\sigma_{-5}(n)\sigma_{-3}(n)\Big[\frac{1}{4y^3} + \frac{n}{4y^2}\Big]\,,\\
                a_0(2,4;4;y)&\notag= \frac{y^6}{3827250} - \frac{y^3\zeta_3}{28350} - \frac{\zeta_7}{720y} - \frac{25\zeta_9}{432y^3} +
                \frac{5\zeta_3\zeta_7}{64y^4}\\*
                &\phantom{=} + \sum_{n=1}^\infty e^{-4ny}\sigma_{-7}(n)\sigma_{-3}(n)\Big[\frac{5}{32y^4}+\frac{5n}{24y^3} +\frac{n^2}{12y^2}\Big]\,.
 \end{align}
 Equation \eqref{eq_examples2k} is identical to the result of \cite{DHoker:2017zhq} for the exponentially suppressed terms of the MGF $C_{2,1,1}$ once we use the fact that $\FFp{2}{2}{2} = -C_{2,1,1} +\frac{9}{10}\EE_{4}$.
 
 In the $(3,k)$ sector, with $k\geq3$, we encounter two choices of eigenvalues $s\in\{k-1,k+1\}$:
 \begin{align}\label{eq_examples3k}
          a_0(3,3;2;y)&= \frac{y^6}{6251175} - \frac{y\zeta_5}{630} - \frac{5\zeta_7}{288y} + \frac{\zeta_5^2}{32y^4} + \sum_{n=1}^\infty e^{-4ny}\sigma_{-5}(n)^2\Big[\frac{1}{16y^4}+\frac{n}{4y^3} + \frac{n^2}{8y^2}\Big]\,,\\
          a_0(3,3;4;y)&= \frac{2y^6}{8037225} - \frac{y\zeta_5}{3780} - \frac{35\zeta_9}{1152y^3} + \frac{9\zeta_5^2}{128y^4}  + \sum_{n=1}^\infty e^{-4ny}\sigma_{-5}(n)^2\Big[\frac{9}{64y^4}+\frac{n}{4y^3} + \frac{n^2}{8y^2}\Big]\,,\\
          a_0(3,4;3;y)&\notag= \frac{y^7}{80372250} - \frac{y^2\zeta_5}{25200} - \frac{\zeta_7}{4536} - \frac{49\zeta_9}{11520y^2} + \frac{5\zeta_5\zeta_7}{256y^5} \\*
          &\phantom{=} + \sum_{n=1}^\infty e^{-4ny}\sigma_{-7}(n)\sigma_{-5}(n)\Big[\frac{5}{128y^5}+\frac{5n}{32y^4} + \frac{7n^2}{48y^3} + \frac{n^3}{24y^2}\Big]\,,\\
          a_0(3,4;5;y)&\notag= \frac{y^7}{49116375} - \frac{y^2\zeta_5}{113400} - \frac{\zeta_7}{15120} - \frac{77\zeta_{11}}{4608y^4} + \frac{3\zeta_5\zeta_7}{64y^5} \\*
          &\phantom{=} + \sum_{n=1}^\infty e^{-4ny}\sigma_{-7}(n)\sigma_{-5}(n)\Big[\frac{3}{32y^5}+\frac{37n}{192y^4} + \frac{7n^2}{48y^3} + \frac{n^3}{24y^2}\Big]\,.
 \end{align}
 As a last example, in the $(4,k)$ sector with $k\geq 4$ and eigenvalues $s\in\{k-2,k,k+2\}$, we have:
 \begin{align}\label{eq_examples4k}
    a_0(4,4;2;y)&\notag= \frac{y^8}{1205583750} - \frac{y\zeta_7}{7560} - \frac{5\zeta_9}{3888y} + \frac{5\zeta_7^2}{512y^6} \\*
    &\phantom{=}+ \sum_{n=1}^\infty e^{-4ny}\sigma_{-7}(n)^2\Big[\frac{5}{256y^6}+\frac{5n}{64y^5} + \frac{35n^2}{288y^4} + \frac{5n^3}{72y^3} + \frac{n^4}{72y^2}\Big]\,,\\
    a_0(4,4;4;y)&\notag= \frac{y^8}{982327500} - \frac{y\zeta_7}{45360} - \frac{7\zeta_{11}}{6912y^3} + \frac{5\zeta_7^2}{384y^6} \\*
    &\phantom{=}+ \sum_{n=1}^\infty e^{-4ny}\sigma_{-7}(n)^2\Big[\frac{5}{192y^6}+\frac{5n}{48y^5} + \frac{25n^2}{192y^4} + \frac{5n^3}{72y^3} + \frac{n^4}{72y^2}\Big]\,,\\
    a_0(4,4;6;y)&\notag= \frac{y^8}{580466250} - \frac{y\zeta_7}{113400} - \frac{5055\zeta_{13}}{530688y^5} + \frac{25\zeta_7^2}{768y^6} \\
    &\phantom{=}+ \sum_{n=1}^\infty e^{-4ny}\sigma_{-7}(n)^2\Big[\frac{25}{384y^6}+\frac{29n}{192y^5} + \frac{7n^2}{48y^4} + \frac{5n^3}{72y^3} + \frac{n^4}{72y^2}\Big]\,.
\end{align}
}
One can check that these results are in agreement with the differential equation~\eqref{int:Fmk}.

\subsection{Exact results}
\label{sec:3.3}

From the general results derived previously, we know that the zero-mode \eqref{eq:0modefinal} for the modular invariant function $\FFp{s}{m}{k}$ is given by
\begin{equation}
a_0(m,k;s;y) = {\rm{P}}_{m,k}^{(s)}(y) +  {\rm{NP}}_{m,k}^{(s)}(y)\,,\label{eq:zeromode}
\end{equation}
with the perturbative terms given by the Laurent polynomials \eqref{eq:FLP} and the non-perturbative terms 
$$ {\rm{NP}}_{m,k}^{(s)}(y) = \lim_{\epsilon \to 0} {\rm{NP}}^{\epsilon}_{\pm}(m,k;s;y)\,,$$
simply obtained from \eqref{eq_F_Stokes} by sending $\epsilon\to 0$.

We can use the general result \eqref{eq_F_Stokes} to write the non-perturbative terms as
\begin{equation}
 {\rm{NP}}_{m,k}^{(s)}(y) = \sum_{n>0} e^{-4ny} \frac{n^{k+m-2} \sigma_{1-2m}(n)\sigma_{1-2k}(n)}{\Gamma(m)\Gamma(k)} \phi_{m,k}^{(s)}(4 n y)\,,\label{eq:NPphi}
\end{equation}
where we used the divisor sum identity $\sigma_s(n) = n^s\sigma_{-s}(n)$, and defined $\phi_{m,k}^{(s)}(y)$ by
\begin{align}
\phi_{m,k}^{(s)}(y) &\label{eq:phiNP}= \lim_{\epsilon \to 0} \Big[ 8 \Gamma(k) y^{1+m-k}\sum_{\ell=k-m+1}^{k-1} \frac{g^+_{m,k,\ell,s}(-1)^{\ell+k}}{\Gamma(\ell)}\\
&\nn \times \int_0^\infty e^{-yt} (t+1)^{m+\ell-k-1}\,t^{k+m+\epsilon-2\ell}\,{}_2\tilde{F}_1(1-\ell,k+m-1;k+m+1+\epsilon-2\ell\vert -t) \,\dd t \Big]\,.
\end{align}

The non-perturbative terms in the zero Fourier mode could have also been obtained by using the ansatz \eqref{eq:NPphi} and substituting it into the inhomogeneous Laplace equation \eqref{int:Fmk} satisfied by the $\FFp{s}{m}{k}$. From the Fourier mode expansion for the Eisenstein series \eqref{eq:FEk}, we can readily isolate the $(q\bar{q})^n$ contribution of the source term ${\rm{E}}_m{\rm{E}}_k$. This results in a second-order differential equation for $\phi_{m,k}^{(s)}(y)$ that could be solved using a Laurent series ansatz. The solution found in this way can be checked to agree with the results presently obtained via resurgent analysis.

From the discussion below \eqref{eq:gamma}, we have that $\phi_{m,k}^{(s)}(y)$ is a polynomial of degree $k+m-2$ in $y^{-1}$ with rational coefficients. 
We will now prove that
\begin{align}
&\label{eq_dom_p}\phi_{m,k}^{(s)}(y) = \\
    &\notag \frac{8}{y^2} + \frac{8[m(m-1)+k(k-1)-4]}{y^3}+4 \frac{\Big\{[m(m-1)+k(k-1)-7]^2 +2 s(s-1) -13
\Big\}}{y^4} +O(y^{-5}).
\end{align}
Note that the coefficients of higher corrections in $y^{-1}$ will in general have a dependence from the eigenvalue $s$.
 
By using the integral transform \eqref{eq:gamma}, we observe that the leading contribution to $\phi^{(s)}_{m,k}(y)$ as $y\to\infty$ comes from the lowest power of $t$ in the integrand of equation \eqref{eq:phiNP}. 
To isolate this monomial, we start by noting that the lowest exponent for the factor $t^{k+m+\epsilon-2\ell}$ is clearly given by the highest value of the parameter $\ell = \ell_{max} = k-1$. In this case we have a simplification for the coefficients \eqref{eq:fkm} $g^+_{m,k,\ell,s}$ appearing in \eqref{eq:phiNP}, in that $g^+_{m,k,k-1,s} = \frac{1}{k-1}$ is independent of the eigenvalue $s$.

To obtain the lowest power of $t$ in the integrand of \eqref{eq:phiNP}, we similarly have to choose the constant term for both the hypergeometric series as well as the binomial when $\ell=k-1$:
\begin{align}
&{}_2\tilde{F}_1(2-k,k+m-1;m+3+\epsilon-k\vert -t) \nn= \frac{1}{\Gamma(m+3+\epsilon-k)}+\frac{(k-2)(k+m-1)}{\Gamma(m+4+\epsilon-k)}t+O\big(t^2\big)\,,\\
&(t+1)^{m-2} \label{eq:smallt}= 1+(m-2) t+O(t^2)\,.
\end{align} 
We then arrive at the leading, large-$y$ asymptotic for \eqref{eq:phiNP} given by
\begin{align}
&\phi_{m,k}^{(s)}(y) \label{dom_contrib_NP}\sim \lim_{\epsilon \to 0} \Big[ 8 \Gamma(k) y^{1+m-k} \frac{g^+_{m,k,k-1,s}}{\Gamma(k-1)} \int_0^\infty e^{-yt} \,\frac{t^{m+2+\epsilon-k}}{\Gamma(m+3+\epsilon-k)} \,\dd t \Big]  \sim \frac{8}{y^2} + \ldots\,,
\end{align}
where we used the standard integral \eqref{eq:gamma} and reproduced the leading order in \eqref{eq_dom_p}.

For the sub-leading correction in  \eqref{eq_dom_p} we need to investigate higher powers of $t$ in the integrand of \eqref{eq:phiNP}. Firstly we observe that decreasing $\ell\to \ell_{max} - 1 = k-2$ increases the power of $t$ by $2$ for the $t^{k+m+\epsilon-2\ell}$ term in the integrand. Hence we deduce that the next sub-leading correction comes again from $\ell = \ell_{max} = k-1$ where we consider instead the linear terms in $t$ for the hypergeometric function and the binomial \eqref{eq:smallt}.  As a result, since the coefficient $g^+_{m,k,k-1,s} = \frac{1}{k-1}$ does not depend from the eigenvalue $s$, we have that, just like for the leading term, the $\frac{1}{y^3}$ coefficient must once more be eigenvalue independent. 

The calculation is very similar to the one presented above
\begin{align}
\phi_{m,k}^{(s)}(y) \label{dom_contrib_NP2}&\nn\sim \frac{8}{y^{2}} + \lim_{\epsilon \to 0} \Big[ 8 \,y^{1+m-k}\int_0^\infty e^{-yt} \,\frac{t^{m+3+\epsilon-k}}{\Gamma(m+3+\epsilon-k)} \Big[(m-2) + \frac{(k-2)(k+m-1)}{m+3+\epsilon-k} \Big]\,\dd t \Big] \\
& \sim \frac{8}{y^2} +\frac{8[m(m-1)+k(k-1)-4]}{y^3} +\ldots \,,
\end{align}
and we reproduce, as anticipated, the sub-leading term of equation \eqref{eq_dom_p}.

Getting analytic expressions for higher-order terms becomes slightly more complicated, since multiple values of $\ell$ in \eqref{eq:phiNP} start contributing and the coefficients $g^+_{m,k,\ell,s}$, see \eqref{eq:fkm}, are in general eigenvalue dependent, thus higher-order terms do depend on the eigenvalue $s$ as well. 
For example, we can repeat a very similar discussion to the one above above for the $O(y^{-4})$ term, which receives two different contributions - one from $\ell = k-1$ and a second one from $\ell = k-2$. By using \eqref{eq:fkm} to obtain the coefficient $g^+_{m,k,k-2,s}$
\begin{equation}
g^+_{m,k,k-2,s} = \frac{m(m-1)}{(k-2)} + \frac{(k-s-1)(k+s-2)}{(k-1)(k-2)}\,,
\end{equation}
 and then collect the appropriate powers of $t$ in the integrand, we arrive at
\begin{align}
\label{dom_contrib_NP3}
&\phi_{m,k}^{(s)}(y) \sim\\
&\nn \frac{8}{y^2} +\frac{8[m(m-1)+k(k-1)-4]}{y^3} +4 \frac{\Big\{[m(m-1)+k(k-1)-7]^2 +2 s(s-1) -13
\Big\}}{y^4} +\ldots\,.
\end{align}

All of the results here discussed can be checked for comparison with the examples given in section~\ref{sec:3.2} and are consistent with the Laplace equation \eqref{int:Fmk}.

\section{Modularity and recovering the small-$y$ behaviour}
\label{sec:4}

Up until now we have used the asymptotic nature of the large-$y$ perturbative expansion to reconstruct the non-perturbative, exponentially suppressed $(q\bar{q})^n$ corrections via resurgent analysis. Now we want to understand a similar, yet conceptually different problem, namely is it possible to reconstruct the perturbative data, i.e. the Laurent polynomials \eqref{eq:FLP}, from the small-$y$ expansion of the $(q\bar{q})^n$ terms? We will see that, complementary to resurgence, modularity will play a crucial role.

First of all, we recall here an important lemma proved in \cite{Green:2014yxa}.
\vspace{0.1cm}

\textbf{Lemma.} \textit{If ${\rm{F}}(\tau)$ is an $\SLtwoZ$ invariant function on the upper half-plane such that at the cusp $y\to\infty$, with $y=\pi \tau_2 = \pi\, \rm{Im}\,\tau$, it satisfies the growth condition ${\rm{F}}(\tau) = O(y^s)$ with $s>1$, then each of its Fourier modes ${\rm{F}}_n(y) = \int_0^1 {\rm{F}}(\tau_1+iy/\pi) e^{-2\pi i n \tau_1} \dd\tau_1$ satisfies the bound ${\rm{F}}_n(y) = O(y^{1-s})$ in the limit $y\to 0$.}
\vspace{0.1cm}

Very roughly, the key idea behind this lemma is that a cuspidal growth of order $y^s$ suggests that the modular invariant function ${\rm F}(\tau)$ must be bounded by ${\rm E}_s(\tau)$ on the whole upper half-plane and since for small $y$ we have ${\rm E}_s(\tau) = O(y^{1-s})$, then the same bound must hold for ${\rm F}(\tau)$.
Let us apply this lemma to our modular invariant functions $\FFp{s}{m}{k}$ and in particular let us try and understand the small-$y$ behaviour of its zero-mode \eqref{eq:zeromode}.

{}From the explicit Laurent polynomial \eqref{eq:FLP} it is clear that $\FFp{s}{m}{k}(\tau) = O(y^{k+m})$ as $\tau\to i\infty$, hence from the lemma we deduce that for small $y$ each Fourier mode of $\FFp{s}{m}{k}$ cannot be more singular than $O(y^{1-k-m})$. We can easily see from \eqref{eq:FLP} that, for the spectrum of eigenvalues considered here, none of the perturbative terms is more singular than $y^{1-k-m}$ and we conclude that the $(q\bar{q})^n$ terms \eqref{eq:NPphi}, which were exponentially suppressed for large $y$, can at most diverge as $y^{1-k-m}$ as $y\to 0 $.

We can run a more refined argument to obtain analytically part of the small-$y$ limit of the $(q\bar{q})^n$ terms.
To this end we can consider the modular invariant linear combination
\begin{equation}\label{eq:lincomb}
{\rm F}(\tau) = \FFp{s}{m}{k}(\tau) + \alpha \,{\rm E}_{m+k}(\tau)\,,
\end{equation}
where the constant $\alpha$, given by
\begin{equation}
\alpha= \frac{ {\rm B}_{2m} {\rm B}_{2k} (2m+2k)! }{ {\rm B}_{2m+2k}(k{+}m{-}s)(k{+}m{+}s{-}1)(2m)! (2k)!},\,
\end{equation}
 is chosen in a such a way (see \eqref{eq:FEk} and \eqref{eq:FLP}) that the coefficient of the leading term $y^{k+m}$ of \eqref{eq:lincomb} is vanishing.

If we assume that $k> m$, we have thus obtained a new auxiliary modular invariant function ${\rm F}(\tau)$ with the tamer
growth at the cusp ${\rm F}(\tau) = O(y^{1+k-m})$. Note that we have excluded the diagonal case, $k=m$, since $F(\tau)$ would grow at the cusp linearly as $O(y^{1})$, hence the Lemma cannot be applied directly; we will however show a diagonal example where the results are consistent with the non-diagonal expectations and we can view the diagonal case as the limit $k\to m$.

By applying the lemma to ${\rm F}(\tau)$ we deduce that its small-$y$ limit cannot be more singular than $O(y^{m-k})$. 
However, if we inspect all the powers appearing in the perturbative expansion \eqref{eq:FEk} and \eqref{eq:FLP}  of the zero-mode, we find that the terms $y^{1-s}, y^{2-k-m}$, coming from $\FFp{s}{m}{k}(\tau)$, and the term $y^{1-k-m}$, coming from $\alpha\, {\rm E}_{k+m}(\tau)$, all violate the bound.
Since the addition of $\alpha\, {\rm E}_{k+m}(\tau)$ does not modify the $(q\bar{q})^n$ sector, we must conclude that the small-$y$ limit of the $(q\bar{q})^n$ terms \eqref{eq:NPphi} must exactly cancel against these singular terms.
The small-$y$ expansion of the $(q\bar{q})^n$ must then take the form:
\begin{align}
 {\rm{NP}}_{m,k}^{(s)}(y) 
 &\nn=- c_{m,k}^{(s)} \zeta_{k+m+s-1} y^{1-s}-  \frac{4^{3-m-k}\Gamma(2m{-}1)\Gamma(2k{-}1)\zeta_{2m{-}1}\zeta_{2k{-}1}}{[\Gamma(m)\Gamma(k)]^2 (k{+}m{-}s{-}1)(k{+}m{+}s{-}2)} y^{2-k-m} \\
 &\label{eq:smally}-\alpha\frac{4(2m+2k{-}3)!  \zeta_{2m+2k-1}}{(m+k{-}2)!(m+k{-}1)!} (4y)^{1-m-k} + O(y^{m-k})\,.
\end{align}

Obtaining this expression directly from the small-$y$ limit of \eqref{eq:NPphi} is not straightforward. A somewhat naive way to proceed is to expand the exponential factor $(q\bar{q})^n=e^{-4 n y}$ for small-$y$ and compute the sum over $n$ term by term via its analytic continuation as a Dirichlet series using Ramanujan's identity \eqref{eq:Rama}.

To illustrate this, we first repeat the calculation, discussed in the previous section, to obtain the most singular term at small-$y$ for $\phi_{m,k}^{(s)}(y)$
\begin{equation}
\phi_{m,k}^{(s)}(y) =\frac{8 \Gamma(2m-1)\Gamma(2k-1)}{(k+m-s-1)(k+m+s-2)\Gamma(m)\Gamma(k)} y^{2-k-m}+ O(y^{3-k-m})\,.
\end{equation}
We can now consider its contribution in the small-$y$ limit to ${\rm{NP}}_{m,k}^{(s)}(y) $ given by:
\begin{align}
 &{\rm{NP}}_{m,k}^{(s)}(y)  = \sum_{n>0} e^{-4ny} \frac{n^{k+m-2} \sigma_{1-2m}(n)\sigma_{1-2k}(n)}{\Gamma(m)\Gamma(k)} \\
 &\nn \qquad\qquad \qquad\times \Big[\frac{8 \Gamma(2m-1)\Gamma(2k-1)}{(k+m-s-1)(k+m+s-2)\Gamma(m)\Gamma(k)} (4 n y)^{2-k-m}+ O(y^{3-k-m})\Big]\\
 &\nn \sim \frac{8 \Gamma(2m-1)\Gamma(2k-1)}{(k+m-s-1)(k+m+s-2)[\Gamma(m)\Gamma(k)]^2} (4 y)^{2-k-m} \sum_{n>0}  \sigma_{1-2m}(n)\sigma_{1-2k}(n)+ O(y^{3-k-m})\\
 &\nn \sim -  \frac{4^{3-m-k}\Gamma(2m{-}1)\Gamma(2k{-}1)\zeta_{2m{-}1}\zeta_{2k{-}1}}{[\Gamma(m)\Gamma(k)]^2 (k{+}m{-}s{-}1)(k{+}m{+}s{-}2)} y^{2-k-m} + O(y^{3-k-m})\,,
\end{align}
where we expanded the exponential term $e^{-4ny}=1+O(y)$ to leading order at small-$y$ and used the analytic continuation at $s=0$ of Ramanujan's identity to resum
\begin{equation}
\sum_{n>0}  \sigma_{1-2m}(n)\sigma_{1-2k}(n) ``=\mbox{''} \zeta(0)\zeta(2m-1)\zeta(2k-1)\,.
\end{equation}
 This calculation reproduces precisely the expected $y^{2-k-m}$ term in equation \eqref{eq:smally} . 
 
 Using the explicit examples \eqref{eq_examples2k}, \eqref{eq_examples3k} and \eqref{eq_examples4k} presented before, it is possible to perform a similar argument to compute also the sub-leading corrections \eqref{eq:smally} by means of analytically continuing the sum over $n$ as a Dirichlet series. We notice, however, that the most singular term in \eqref{eq:smally} is of order $y^{1-m-k}$ and cannot possibly be obtained via this na\"ive analysis. 

A more careful analysis of the small-$y$ expansion of \eqref{eq:NPphi} can be derived from a Mellin transform argument.
From the generic expression \eqref{eq:NPphi}, it is easy to see that ${\rm{NP}}^{(s)}_{m,k}(y)$ is given by a finite linear combination of functions defined by 
\begin{equation}
    D_{a,b;c}(y) = \sum_{n=1}^\infty \frac{\sigma_a(n)\sigma_b(n)}{n^c}e^{-ny},\label{eq:Dabcmain}
\end{equation}
with $a=1-2m,b=1-2k$ and $c\in \mathbb{Z}_{\leq 0}$.

In appendix \ref{app:small_y} we derive the small-$y$ behaviour \eqref{eq:Genabc} of the function $D_{a,b;c}(y)$ with $a,b,c\in\mathbb{C}$ generic.
Using Mellin inversion formula, the asymptotic expansion at $y \to 0 $ of $ D_{a,b;c}(y)$ is related to the poles and residues of its Mellin transform $M_{a,b;c}(y)$.

We refer to appendix \ref{app:small_y} for the general discussion and present here a few concrete examples.
Let us consider the non-perturbative terms ${\rm{NP}}^{(3)}_{2,3}(y)$ for the zero Fourier mode of the depth-$2$ modular function $\FFp{3}{2}{3}$, which are given by \eqref{eq_examples2k_2}
\begin{align*}
    {\rm{NP}}^{(3)}_{2,3}(y) &= \frac{1}{4}\sum_{n=1}^\infty\sigma_{-5}(n)\sigma_{-3}(n)e^{-4ny}\Big[\frac{1}{y^3}+\frac{n}{y^2}\Big] \\
    &= \frac{1}{4y^3}D_{-5,-3;0}(4y)+\frac{1}{4y^2}D_{-5,-3;-1}(4y).
\end{align*}
The relevant Mellin transforms, see \eqref{M_formula}, are
\begin{align}
        M_{-5,-3;0}(t) &= \int_0^\infty D_{-5,-3;0}(y)\,y^{t-1}\dd y= \frac{\Gamma(t)\zeta(t)\zeta(3+t)\zeta(5+t)\zeta(8+t)}{\zeta(8+2t)}\,,\\
        M_{-5,-3;-1}(t) &= \int_0^\infty D_{-5,-3;-1}(y)\,y^{t-1}\dd y= \frac{\Gamma(t)\zeta(-1+t)\zeta(2+t)\zeta(4+t)\zeta(7+t)}{\zeta(6+2t)}\,,
\end{align}
from which it is easy to see that $M_{-5,-3;0}(t)$ has simple poles at $t\in \mathbb{Z}$ in the range $-8\leq t\leq 1$, while $M_{-5,-3;-1}(t)$ has simple poles at $t\in\mathbb{Z}$ in the range $-7\leq t\leq 2$, excluding $t=1$.

Referring to appendix \ref{app:small_y} for the details, we can use the Mellin inversion formula \eqref{D_int} and, after the little exercise of computing the residues at these poles, we arrive at the small-$y$ expansion for $ {\rm{NP}}^{(3)}_{2,3}(y)$:
\begin{multline}\label{eq_NP233}
    {\rm{NP}}^{(3)}_{2,3}(y) \sim \frac{11\zeta_9}{128y^4}-\frac{\zeta_3\zeta_5}{8y^3}+\frac{7\zeta_7}{64y^2}-\frac{\zeta_3^2}{42y}+\frac{\zeta_5}{360}+\frac{\zeta_3y^2}{1890}-\frac{\zeta_7y^3}{3240\zeta_5}+\frac{\zeta_3\zeta_5y^4}{23625\zeta_7}-\frac{y^5}{297675}\,.
\end{multline}
A comparison with (\ref{eq_examples2k_2}) reveals that the small-$y$ limit of the non-perturbative terms not only matches perfectly the expected behaviour \eqref{eq:smally} but it actually cancels exactly the full Laurent polynomial part:
\begin{equation}
    {\rm{NP}}^{(3)}_{2,3}(y) \sim \frac{11\zeta_9}{128y^4}-P^{(3)}_{2,3}(y) 
    -\frac{\zeta_3^2}{42y}
    -\frac{\zeta_7y^3}{3240\zeta_5}+\frac{\zeta_3\zeta_5y^4}{23625\zeta_7}\,.
\end{equation} 
The difference between the small-$y$ limit of the non-perturbative sector and the Laurent polynomial is given by the expected $y^{1-k-m}$ monomial of \eqref{eq:smally} and terms vanishing as $y\to0$. Although such terms present novel type of coefficients, in the form of ratios of zeta values, they do respect uniform transcendentality with standard weight assignment.

As a second example, we can analyse the small-$y$ limit of  ${\rm{NP}}^{(s)}_{m,m}(y)$, i.e. the non-perturbative terms in the diagonal sector $k=m$. 
For simplicity let us consider  ${\rm{NP}}^{(2)}_{2,2}(y)$, i.e. the non-perturbative terms of the modular function $\FFp{2}{2}{2}$ that are given by \eqref{eq_examples2k}
\beq
{\rm{NP}}^{(2)}_{2,2}(y) = \sum_{n=1}^\infty \frac{e^{-4ny}\sigma_{-3}(n)^2}{2y^2} = \frac{1}{2y^2}D_{-3,-3;0}(4y).
\eeq
The corresponding Mellin transform is given by
\beq
M_{-3,-3;0}(t) = \int_0^\infty D_{-3,-3;0}(y)\,y^{t-1}\dd y = \frac{\Gamma(t)\zeta(t)\zeta(3+t)^2\zeta(6+t)}{\zeta(6+2t)},
\eeq
which has poles at  $t\in\mathbb{Z}$ in the range $ -6\leq t\leq 1$.
The key difference between the diagonal sector and the previous, non-diagonal example is the appearance of a double pole at $t=-2$, while all others are simple poles.
Referring again to appendix \ref{app:small_y} for all the details, in this case  we have that a second order pole in the Mellin transform signals the presence of logarithmic corrections, $\log y$, in the asymptotic expansion as $y\to0$ of ${\rm{NP}}^{(2)}_{2,2}(y)$.

The asymptotic expansion as $y\to 0$ of ${\rm{NP}}^{(2)}_{2,2}(y)$ is given by
\begin{align}
{\rm{NP}}^{(2)}_{2,2}(y) &\nn \sim \frac{7\zeta_7}{48y^3} - \frac{\zeta_3^2}{4y^2} + \frac{5\zeta_5}{12y} + \frac{\zeta_3}{15}\Big[ \log{\Big(\frac{8\pi y}{A^{24}}\Big)}+\frac{\zeta'_3}{\zeta_3}-\frac{\zeta'_4}{\zeta_4}\Big] +\frac{\zeta_3y}{45} - \frac{\zeta_5y^2}{108\zeta_3} +\frac{2\zeta_3^2y^3}{2835\zeta_5} - \frac{y^4}{20250}\\
&\label{eq_NP_222}\sim \frac{7\zeta_7}{48y^3} - P^{(2)}_{2,2}(y)  +  \frac{\zeta_3}{15}\Big[ \log{\Big(\frac{8\pi y}{A^{24}}\Big)}+\frac{\zeta'_3}{\zeta_3}-\frac{\zeta'_4}{\zeta_4}\Big]  - \frac{\zeta_5y^2}{108\zeta_3} +\frac{2\zeta_3^2y^3}{2835\zeta_5}\,,
\end{align}
where $A$ is the Glaisher–Kinkelin constant, $\log A = \frac{1}{12}-\zeta'(-1)$.
 If we assign transcendental weight 1 to $\log{(8\pi y/A^{24})}+\zeta'_3/\zeta_3 -\zeta'_4/\zeta_4$, then the result respects uniform transcendentality. This appears to be a variant of the transcendentality assignments in~\cite{DHoker:2019blr}.
 Note that we reproduce again the expected behaviour \eqref{eq:smally}, furthermore, after comparison with the Laurent polynomial $P^{(2)}_{2,2}(y)$ (\ref{eq_examples2k}), we have the stronger statement that the non-perturbative terms cancel exactly the perturbative ones as for the previous non-diagonal example.

For the general small-$y$ expansion of ${\rm{NP}}^{(s)}_{m,k}(y)$, we see that \eqref{eq:NPphi} can be expressed as a finite linear combination of building blocks $D_{a,b;c}(y)$, defined in \eqref{eq:Dabcmain}, with $a=1-2m,b=1-2k$ and $c\in \mathbb{Z}_{\leq 0}$. For this range of parameters, it is easy to see from the general formula \eqref{eq:Genabc}, derived in appendix \ref{app:small_y}, that the $y$-coefficients appearing the small-$y$ limit ${\rm{NP}}^{(s)}_{m,k}(y)$ can be at most ratios of bilinears in odd zetas divided by a single odd zeta, or in the diagonal case $m=k$ contain at most one derivative of a Riemann zeta.

Although quite different in spirit from the main message of this paper, the small-$y$ behaviour can also be retrieved by exploiting the spectral decomposition of the modular functions $\FFp{s}{m}{k}$ in terms of ${\rm L}^2$-normalisable eigenfunctions of the ${\rm{SL} }(2,\mathbb{R})$ invariant Laplacian $\Delta = 4(\Im\tau)^2\partial_\tau\partial_{\bar{\tau}}$. 
An interesting question is understanding the interplay between resurgence theory and the spectral analysis of $\FFpm{s}{m}{k}$, see \cite{Klinger:2018,Collier:2022emf} for the spectral decomposition of $\FFp{s}{m}{k}$ with $s,m,k\in\mathbb{C}$. In particular, amongst the ${\rm L}^2$-normalisable eigenfunctions of the Laplacian, \textit{non-holomorphic} cusp forms should play a special role in reconstructing the ``instanton'' sector, i.e. $q^n$ terms in the Fourier decomposition of the modular functions $\FFpm{s}{m}{k}$. Previous works \cite{Dorigoni:2021ngn} have shown from the different point of iterated integrals that \textit{holomorphic} cusp forms do also play a role in the instantonic sector, however there is no obvious or straightforward connection between the holomorphic and non- cusp functions. We aim to address these issues in future works.
 
As a concluding remark for this section we want to stress that if resurgent analysis allows us to retrieve the exponentially suppressed and non-perturbative corrections at large-$y$ from the perturbative data, modularity dramatically intertwines the two and permits us to reconstruct the Laurent polynomial from the small-$y$ behaviour of the infinite tower of $(q\bar{q})^n$ terms, no longer exponentially suppressed.

\section{Conclusions}
\label{sec:Conc}

In this paper, we have studied the non-perturbative terms of the form $(q\bar{q})^n$ that enter in the zero Fourier mode of the modular functions $\FFp{s}{m}{k}$ satisfying the inhomogeneous Laplace equation~\eqref{int:Fmk}. The way we obtained these non-perturbative terms was by using Cheshire cat resurgence and we checked that the results obtained in this way are consistent with the differential equation. Combining the resurgent analysis results with modularity we could also constrain the small-$y$ limit and perform further consistency checks. The focus in this paper was on depth-two functions and we remark that Laplace systems similar to~\eqref{int:Fmk}  have been recently investigated in~\cite{Drewitt:2021} at depth three.

In \cite{Dorigoni:2021jfr,Dorigoni:2021ngn}, the non-zero modes, $a_n(m,k; s; y)$, for the $\tau_1 =\mbox{Re}\,\tau$ Fourier decomposition of \eqref{eq:FourierExp} of $\FFp{s}{m}{k}$ were discussed in detail. 
The general $a_n(m,k; s; y)$ has a form very similar to the zero-mode \eqref{eq:zeromode} discussed in this work, namely an overall exponentially suppressed factor $e^{-2|n|y}$, i.e. what one would call the real part of the instanton action, multiplied by a truncating Laurent polynomial, i.e. the would-be perturbative expansion in the $n$-instanton sector. Furthermore, just like in \eqref{eq:zeromode}--\eqref{eq:NPphi}, the truncating Laurent perturbative part receives an infinite tower of exponentially suppressed corrections, $e^{-4Ny}$ with $N>0$, corresponding to $(q\bar{q})^N$ contributions, multiplied by a truncating Laurent polynomial similar to \eqref{eq:NPphi}.

As shown in \cite{Dorigoni:2021jfr,Dorigoni:2021ngn}, while the zero-mode perturbative sector is given by a Laurent polynomial \eqref{eq:FLP} in $y$ with coefficients given by rational numbers times at most bilinears in odd zeta values, the perturbative expansion in the non-zero mode sector does involve a larger class of numbers including completed $L$-values of holomorphic cusp forms. 

On the other hand, just like in the present case \eqref{eq:phiNP}, the $(q\bar{q})^N$ terms in the non-zero mode sector is given by a Laurent polynomial in $y$ with rational coefficients. It would be extremely interesting to firstly understand how to generalise the zero-mode result \eqref{eq:Iabr} from \cite{Dorigoni:2019yoq} and obtain the perturbative expansion in the non-zero mode sector where presumably the general Kloosterman sum in \eqref{eq:nonzeromode} will be the source for the aforementioned completed $L$-values.

Secondly, very much in the spirit of the present work, one should be able to retrieve the $(q\bar{q})^N$ terms in the non-zero mode sector from Cheshire cat resurgence applied to a suitable deformation of the perturbative expansion in the non-zero mode sector. 

The importance of these questions is even more prominent when considering a second class of modular invariant functions, denoted by $\FFm{s}{m}{k}$ in \cite{Dorigoni:2021jfr,Dorigoni:2021ngn}. Since the $\FFm{s}{m}{k}$ are odd under the involution $\tau \rightarrow - \bar \tau$ of the upper half-plane, they must be cuspidal i.e. they have a vanishing zero Fourier mode $a_0(m,k; s; y)=0$. Although the Poincar\'e seeds for $\FFm{s}{m}{k}$ are completely understood \cite{Dorigoni:2021jfr}, at the present time it is not known how to reconstruct the non-zero Fourier modes, $a_n(m,k; s; y)$, for $\FFm{s}{m}{k}$ via the Poincar\'e series approach reviewed in appendix \ref{app:Poincare}. We believe that a story closely resembling the present work should hold for the non-zero modes of the cuspidal functions $\FFm{s}{m}{k}$ as well.

Furthermore, the importance of our work extends to an even broader class of functions.
Since an equation very similar to~\eqref{int:Fmk} has appeared also in the context of higher-derivative corrections, in particular the schematic term $D^6R^4$, to the type IIB low-energy effective action, where now $\SLtwoZ$ plays the role of U-duality acting on the axio-dilaton~\cite{Green:2005ba}. Even though the inhomogeneous equation looks very similar, an important difference is that in the context of the $D^6R^4$ term the indices $m$ and $k$ on the Eisenstein series in~\eqref{int:Fmk} are half-integral. This leads to $(q\bar{q})^n$ that are multiplied by non-truncating asymptotic series $\phi_{m,k}^{(s)}$ in~\eqref{eq:NPphi} whereas for MGFs only Laurent polynomials entered, see section~\ref{sec:3.2} for examples. This makes the study of the MGFs more tractable. However, the half-integer cases have also shown up in many contexts and we refer to~\cite{Green:1997tv,Green:1998by,Green:2005ba,Green:2014yxa,Pioline:2015yea,Bossard:2015uga,Bossard:2020xod,Chester:2020vyz,Green:2014yxa,Klinger:2018,Chester:2020vyz,Minprogress,FKinprogress} for related work. 

The U-duality-invariant higher derivative couplings are also relevant for precisions tests of the AdS/CFT correspondence.
On the AdS/CFT side, recent developments on the flat-space limit of type-IIB effective actions on AdS$_5 \times S^5$ involving localisation and conformal-bootstrap methods include \cite{Binder:2019jwn, Binder:2019mpb, Chester:2019pvm, Chester:2019jas, Chester:2020vyz, Green:2020eyj,Dorigoni:2021bvj,Dorigoni:2021guq}, and the interplay with correlation functions in ${\cal N}=4$ super Yang--Mills has for instance been investigated in \cite{Okuda:2010ym, Penedones:2010ue, Alday:2018pdi, Alday:2018kkw, Drummond:2019odu, Aprile:2019rep, Drummond:2019hel, Bissi:2020wtv, Bissi:2020woe,Dorigoni:2021rdo,Dorigoni:2022zcr}, see also the recent review \cite{Dorigoni:2022iem}. We expect some form of Cheshire cat resurgence to be also applicable to the cases of higher derivative couplings, where the modular parameter $\tau$ is now representing the axio-dilaton, while the recent work \cite{Hatsdua:2022enx} discusses some resurgent analysis properties of the large-$N$ expansion from the point of view of (integrated) correlation functions in ${\cal N}=4$ super Yang--Mills.

Finally, a key point to understand is the generalisation of our results to higher genus world-sheets.
In a certain degeneration limit, a genus-two world-sheet can be thought of as a genus-one world-sheet where we have ``shrunk'' one of the two handles down to a pair of marked points on the surviving torus. The low-energy expansion of genus-two string scattering amplitudes in the degeneration limit produces then a new class of functions, called elliptic modular graph forms (eMGFs) \cite{DHoker:2020hlp,Hidding:2022vjf}, depending both from the modular parameter $\tau$, as well as a second parameter $z$, encoding the details of the degeneration limit. It is still not fully understood how to compute in general the expansion of eMGFs at the cusp $\tau\to i \infty$ since both the perturbative, Laurent polynomial part, and the non-perturbative corrections do now depend from the additional parameter $z$. We believe a parametric resurgent analysis approach \cite{Costin2008,Crew:2022qoe} (i.e. resurgent analysis with respect to $\tau$, while $z$ remains a parameter) to this problem should allow us to reconstruct the non-perturbative corrections, with their $z$-dependence, from the $z$-dependent perturbative asymptotic expansion at the cusp.


\medskip
\subsection*{Acknowledgements}

We wish to thank Oliver Schlotterer for useful discussions and comments on an earlier version of this work. The authors would like to thank the Isaac Newton Institute of Mathematical Sciences for support and hospitality during the programme ``New connections in number theory and physics'' when work on this paper was undertaken. This work was supported by EPSRC Grant Number EP/R014604/1.

\appendix

\section{Fourier expansions of  Poincar\'e series}
\label{app:Poincare}

Here, we collect some standard results (see e.g.~\cite{Iwaniec:2002}) on relating the Fourier series of a Poincar\'e series to that of its seed, following the notation of~\cite{Dorigoni:2019yoq,Dorigoni:2021jfr}.
We start from the relation
\begin{align}
\summ(\tau) &=   \PS \seeed(\gamma \tau)\,,
\end{align}
between the Poincar\'e series $\summ(\tau)$ and its seed $\seeed(\tau)$ that have the respective Fourier series
\begin{align}
\summ(\tau)=\sum_{\ell\in\mathbb{Z}} a_\ell(\tau_2) e^{2\pi i \ell \tau_1}\,,\quad \quad 
\seeed(\tau) &= \sum_{\ell\in\mathbb{Z}} c_\ell(\tau_2) e^{2\pi i \ell \tau_1}\,,
\end{align}
with $\tau_1=\Re \tau$ and $\tau_2 = \Im \tau $.
The relation between the Fourier coefficients $a_\ell(\tau_2)$ and $c_\ell(\tau_2)$ is given by~\cite{Iwaniec:2002,Fleig:2015vky}:
\begin{align}
a_\ell(\tau_2) &=\label{eq:nonzeromode} c_\ell(\tau_2) + \sum_{d=1}^\infty\sum_{n\in\mathbb{Z}} S(n,\ell;d) \int_{\mathbb{R}} e^{-2\pi i \ell \omega -2\pi i n \frac{\omega}{d^2 (\tau_2^2+\omega^2)}} c_n\Big(\frac{\tau_2}{d^2(\tau_2^2+\omega^2)}\Big)\dd \omega\,.
\end{align}
In the above formula, $S(n,\ell;d)$ denotes the Kloosterman sum
\begin{equation}
\label{eq:Kloos}
S(n,\ell;d) = \sum_{r\in (\mathbb{Z}/d\mathbb{Z})^\times} e^{2\pi i (n r + \ell r^{-1}) / d}\,,
\end{equation}
where $r\in (\mathbb{Z}/d\mathbb{Z})^\times$ denotes the finite sum over all $0\leq r <d$ that are coprime with $d$. If $r$ is coprime with $d$ it has a multiplicative inverse, denoted by $r^{-1}$, in $ (\mathbb{Z}/d\mathbb{Z})^\times$.

The main focus for us is the Fourier zero mode $a_0(\tau_2)$ for which~\eqref{eq:nonzeromode} specialises to
\begin{align}
a_0(\tau_2) &= c_0(\tau_2) + \sum_{d=1}^\infty \sum_{n\in\mathbb{Z}} \sum_{r\in (\mathbb{Z}/d\mathbb{Z})^\times} e^{2\pi i n r/d} \int_{\mathbb{R}} e^{-2\pi i n \frac{\omega}{d^2 (\tau_2^2+\omega^2)}} c_n\Big(\frac{\tau_2}{d^2(\tau_2^2+\omega^2)}\Big)\dd \omega\nn\\
&= I_0(\tau_2)+ I(\tau_2) \,.
\end{align}
As indicated in the second line it is useful to separate the contributions of $c_0$ from those of the $c_n$ with $n\neq 0$, where we defined (changing also variables according to $\omega = \tau_2\,t$)
\begin{align}
\label{eq:Kloost1}
I_0(\tau_2) &= c_0(\tau_2) +\tau_2 \sum_{d=1}^\infty \sum_{r\in (\mathbb{Z}/d\mathbb{Z})^\times} \int_{\mathbb{R}} c_0\Big(\frac{1}{\tau_2 d^2(1+t^2)}\Big)\dd t \,, \nn\\
I (\tau_2) &= \tau_2  \sum_{d=1}^\infty \sum_{n\neq0} \sum_{r\in (\mathbb{Z}/d\mathbb{Z})^\times} e^{2\pi i n r/d} \int_{\mathbb{R}} e^{-2\pi n \frac{i t}{\tau_2 d^2 (1+t^2)}} c_n\Big(\frac{1}{\tau_2 d^2(1+t^2)}\Big)\dd t\,.
\end{align}

In this work only Poincar\'e seeds of a restricted functional form appear. More precisely, all seeds here considered are given by (finite) linear combination of the basic objects
\begin{subequations}
\begin{align}
c_0(y) &\label{eq:Seed0Gen}=  (\pi \tau_2)^r = y^r\,,\\
c_\ell (y) & \label{eq:SeedNZGen}=   \sigma_{a}(\vert \ell \vert) (4\pi \vert \ell \vert )^b \tau_2 ^r e^{-2\pi \vert \ell \vert \tau_2}= \sigma_{a}(\vert \ell \vert) (4\pi \vert \ell \vert )^b (y/\pi)^r e^{-2 \vert \ell \vert y}\,,
\end{align}
\end{subequations}
with $a,b,r\in \CC$ and $y =\pi \tau_2 $. Their contributions to the Laurent polynomial in $a_0(\tau)$ were found in~\cite{Dorigoni:2019yoq} to be
\begin{subequations}
\label{eq:I0I}
\begin{align}
I_0(r;y) &\label{eq:I0}=  y^r + 
\frac{ (-16)^{1 - r} ( 2 r)! ( 2 r {-}3)!}{
{\rm B}_{2r}  (r {-} 2)! (r {-} 1)!} \zeta(2r{-}1) y^{1 - r}\,, \\
I(a,b,r;y) &=  \frac{2^{3-2r+2b}\pi }{\Gamma(r)} \Big(\frac{y}{\pi}\Big)^{1+b-r}\Bigg[ \frac{y}{\pi^2} \frac{\Gamma(b{+}1)\Gamma(2r{-}b{-}2)}{\Gamma(r{-}b{-}1)} \frac{\zeta(2r{-}a{-}2b{-}2)\zeta(1{-}a)}{\zeta(2r{-}a{-}2b{-}1)} \nn\\
&\hspace{2mm} + \left(\frac{y}{\pi^2}\right)^{a+1} \frac{\Gamma(a{+}b{+}1)\Gamma(2r{-}a{-}b{-}2)}{\Gamma(r{-}a{-}b{-}1)}\frac{\zeta(2r {-}a{-}2b{-}2)\zeta(a{+}1)}{\zeta(2r{-}a{-}2b{-}1)} \nn\\
&\hspace{2mm} +\left(\frac{\pi^2}{y}\right)^b 
\sum_{n \geq 0} \left(\frac{-\pi^2}{y}\right)^{n} \frac{\Gamma(2r{+}n{-}1)}{n! \cdot \Gamma(r{+}n)}  \label{eq:Iabr}\\
&\hspace{20mm}\times \frac{\zeta({-}b{-}n)\zeta({-}a{-}b{-}n)\zeta(2r{-}a{-}b{+}n{-}1)\zeta(2r{-}b{+}n{-}1)}{\zeta(2r{+}2n)\zeta(2r{-}a{-}2b{-}1)} \bigg]\,, \notag
\end{align}
\end{subequations}
In view of~\eqref{eq:PSEk}, the result $I_0$ for a seed $c_0(y) = y^r$ is proportional to that of the standard non-holomorphic Eisenstein series $\EE_r$.

The above results~\eqref{eq:I0I} were obtained originally in the range of parameters where the integrals and series converge. We shall also require their values at analytically continued parameter values and refer the reader to~\cite{Dorigoni:2019yoq,Dorigoni:2021jfr} for details on these analytic continuations.

\section{A Mellin transform Lemma}\label{app:small_y}

In this appendix we will derive the asymptotic expansion near $y\to0$ of the series:
\begin{equation}\label{series_def}
    D_{a,b;c}(y) = \sum_{n=1}^\infty \frac{\sigma_a(n)\sigma_b(n)}{n^c}e^{-ny},
\end{equation}
with $a,b,c\in\mathbb{C}$ while $y>0$.
Notice that this series is absolutely convergent for any $y>0$ since $|\sigma_a(n)\sigma_b(n)n^{-c}|\leq n^{2+|a|+|b|+|c|} $. Furthermore, using $\sigma_{a}(n) = n^a \sigma_{-a}(n)$ we have 
$$D_{a,b;c}(y)  = D_{-a,b;c-a}(y) =D_{a,-b;c-b}(y) =D_{-a,-b;c-a-b}(y)\,.$$

To proceed, we wish to evaluate the Mellin transform
\begin{equation}
   M_{a,b;c}(t) \coloneqq  \mathcal{M}[D_{a,b;c}](t) = \int_0^\infty D_{a,b;c}(y)y^{t-1} \dd y.
\end{equation}
Since the series \eqref{series_def} is exponentially suppressed as $y\to\infty$, we conclude that for sufficiently large $\Re(t)>t_0$ the integral converges absolutely. Hence, when $\Re(t)>t_0$, we can commute the sum with the integral and integrate term by term. After using the standard Ramanujan identity (\ref{eq:Rama}) we obtain
\begin{equation}\label{M_formula}
    M_{a,b;c}(t) = \frac{\Gamma(t)\zeta(t+c)\zeta(t+c-a)\zeta(t+c-b)\zeta(t+c-a-b)}{\zeta(2t+2c-a-b)}.
\end{equation}
Although we derived this equation working in the wedge $\Re(t)>t_0$, we have that \eqref{M_formula} is actually the unique meromorphic extension of $M_{a,b;c}(t)$ to the whole complex plane $t\in\mathbb{C}$. 

\begin{figure}
    \centering
    \includegraphics[scale=0.5]{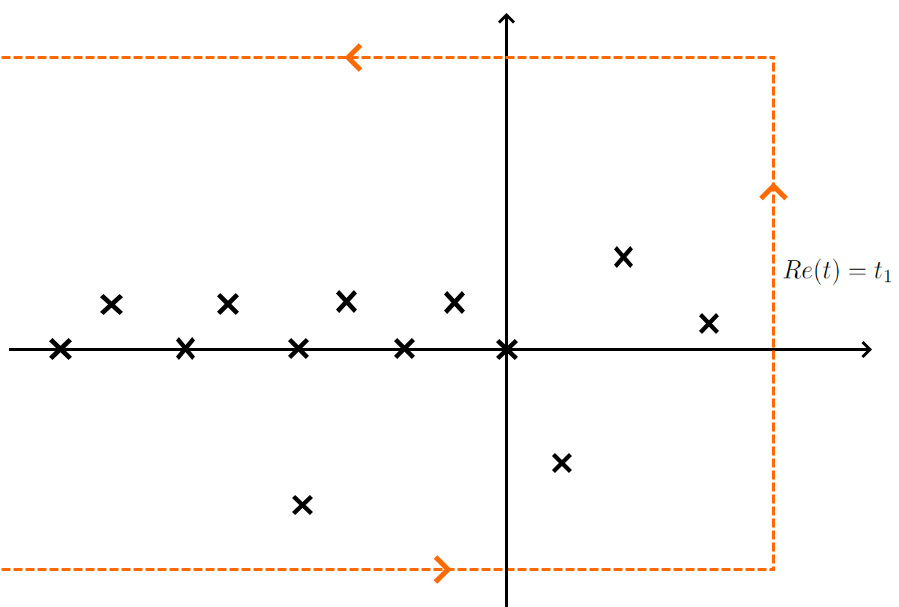}
    \caption{\label{fig:Mellin}\textit{Deformed $t$-integration contour for \eqref{D_int} and poles structure for the Mellin transform $M_{a,b;c}(t)$. For $a,b,c\in\mathbb{C}$ generic, there are two infinite families of poles: $t=-n\,,\,n\in \mathbb{Z}_{\geq 0}$ and $t= -n-c+\frac{a+b}{2}\,,\, n\in\mathbb{Z}_{>0}$, as well as four isolated poles for $t \in\{ 1-c,1+a-c,1+b-c,1+a+b-c\}$.}}

\end{figure}

The asymptotic expansion as $y\to 0$ of $D_{a,b;c}(y)$ is uniquely fixed by the singularities in $t$ of its Mellin transform $M_{a,b;c}(t)$. 
To make this more precise we consider Mellin inversion formula
\begin{equation}\label{D_int}
    D_{a,b;c}(y) = \mathcal{M}^{-1}[M_{a,b;c}](y) = \frac{1}{2\pi i} \int_{t_1-i\infty}^{t_1+i\infty}M_{a,b;c}(t)y^{-t}\dd t,
\end{equation}
where $t_1>t_0$ is arbitrary. The asymptotic expansion as $y\to 0$ of \eqref{D_int} can now be  computed by closing the contour into a loop with $\mbox{Re}\,t<0$, as depicted in figure \ref{fig:Mellin}, and evaluating it using Cauchy's residue theorem (and discarding exponentially suppressed corrections $e^{-4\pi^2/y}$).

Note that the Mellin transform (\ref{M_formula}) is a meromorphic function with an infinite number of poles for generic values of $a,b,c$. Hence from \eqref{D_int}, we expect the expansion for $y\to 0 $ of $D_{a,b;c}(y)$ to be a non-truncating asymptotic expansion. It is easy to see that for generic $a,b,c\in\mathbb{C}$ the Mellin transform (\ref{M_formula}) has only simple poles at locations:
\begin{itemize}
\item $t=-n\,,\,n\in \mathbb{Z}_{\geq 0}$, from the gamma function in the numerator;
\item $t \in\{ 1-c,1+a-c,1+b-c,1+a+b-c\}$, from the zeta functions in the numerator;
\item $t= -n-c+\frac{a+b}{2}\,,\, n\in\mathbb{Z}_{>0}$, from the zeta function in the denominator.
\end{itemize}

Nevertheless, for non-generic values of the parameters $a,b,c$, we can have that zeros at negative even integers of the zeta functions in the numerator, or the pole at one of the zeta function in the denominator, can cancel against the poles listed above thus leaving only a finite number of perturbative terms.
Note that for non-generic values of $a,b,c$ it is also possible to generate higher order poles, thus leading to logarithmic terms, $\log y$, in the asymptotic expansion as $y\to 0$ of $D_{a,b;c}(y)$.

Assuming generic $a,b,c\in\mathbb{C}$, we can compute \eqref{D_int} via residues calculus deforming the contour of integration as depicted in Figure \ref{fig:Mellin} and derive the asymptotic expansion as $y\to 0$ of $D_{a,b;c}(y)$ given by:
\begin{align}
&\label{eq:Genabc} D_{a,b;c}(y) \sim\\
&\nn y^{c-1} \frac{\Gamma(1-c)\zeta(1-a)\zeta(1-b)\zeta(1-a-b)}{\zeta(2-a-b)}+y^{c-a-1} \frac{\Gamma(1+a-c)\zeta(1+a)\zeta(1-b)\zeta(1+a-b)}{\zeta(2+a-b)}\\
&\nn+
y^{c-b-1} \frac{\Gamma(1+b-c)\zeta(1-a)\zeta(1+b)\zeta(1-a+b)}{\zeta(2-a+b)}\\
&\nn+
y^{c-a-b-1}\frac{\Gamma(1+a+b-c)\zeta(1+a)\zeta(1+b)\zeta(1+a+b)}{\zeta(2+a+b)}\\
&\nn +\sum_{n=0}^\infty (-y)^n \frac{\zeta(c-n)\zeta(c-a-n)\zeta(c-b-n)\zeta(c-a-b-n)}{n! \zeta(2c-2n-a-b)}\\
&\nn + \sum_{n=1}^\infty y^{c-\frac{a+b}{2}} (-4\pi^2 y)^n  \frac{\Gamma(\frac{a+b}{2}-c-n)\zeta(\frac{a+b}{2}-n)\zeta(\frac{b-a}{2}-n)\zeta(\frac{a-b}{2}-n) \zeta(-\frac{a+b}{2}-n)}{(2n)! \zeta(2n+1)}\,.
\end{align}

Note that for generic $a,b,c\in\mathbb{C}$, both sums over $n$ are asymptotic, factorially growing series. An interesting exercise would be to use resurgent analysis to derive the exponentially suppressed corrections $e^{-4\pi^2 k/y}$, with $k\in \mathbb{Z}_{>0}$, from a median resummation of such series. As a consistent, full-circle analysis the large-$y$ expansion of such small-$y$ non-perturbative corrections must reproduce back the starting large-$y$ exponentially suppressed series \eqref{series_def}.

\bibliography{cites}

\providecommand{\href}[2]{#2}\begingroup\raggedright\begin{thebibliography}{10}

\bibitem{Green:1999pv}
M.~B. Green and P.~Vanhove, ``{The Low-energy expansion of the one loop type II
  superstring amplitude},''
  \href{http://dx.doi.org/10.1103/PhysRevD.61.104011}{{\em Phys.Rev.} {\bf D61}
  (2000)  104011},
\href{http://arxiv.org/abs/hep-th/9910056}{{\tt arXiv:hep-th/9910056
  [hep-th]}}.

\bibitem{Green:2008uj}
M.~B. Green, J.~G. Russo, and P.~Vanhove, ``{Low energy expansion of the
  four-particle genus-one amplitude in type II superstring theory},''
  \href{http://dx.doi.org/10.1088/1126-6708/2008/02/020}{{\em JHEP} {\bf 02}
  (2008)  020},
\href{http://arxiv.org/abs/0801.0322}{{\tt arXiv:0801.0322 [hep-th]}}.

\bibitem{DHoker:2015gmr}
E.~D'Hoker, M.~B. Green, and P.~Vanhove, ``{On the modular structure of the
  genus-one Type II superstring low energy expansion},''
  \href{http://dx.doi.org/10.1007/JHEP08(2015)041}{{\em JHEP} {\bf 08} (2015)
  041},
\href{http://arxiv.org/abs/1502.06698}{{\tt arXiv:1502.06698 [hep-th]}}.

\bibitem{DHoker:2015wxz}
E.~D'Hoker, M.~B. Green, {\"O}.~G{\"u}rdo{\u{g}}an, and P.~Vanhove, ``Modular
  graph functions,'' \href{http://dx.doi.org/10.4310/CNTP.2017.v11.n1.a4}{{\em
  Commun. Num. Theor. Phys.} {\bf 11} (2017) no.~1, 165--218},
\href{http://arxiv.org/abs/1512.06779}{{\tt arXiv:1512.06779 [hep-th]}}.

\bibitem{DHoker:2016mwo}
E.~D'Hoker and M.~B. Green, ``Identities between modular graph forms,''
  \href{http://dx.doi.org/10.1016/j.jnt.2017.11.015}{{\em J. Number Theory}
  {\bf 189} (2018)  25--80},
\href{http://arxiv.org/abs/1603.00839}{{\tt arXiv:1603.00839 [hep-th]}}.

\bibitem{Green:2013bza}
M.~B. Green, C.~R. Mafra, and O.~Schlotterer, ``{Multiparticle one-loop
  amplitudes and S-duality in closed superstring theory},''
  \href{http://dx.doi.org/10.1007/JHEP10(2013)188}{{\em JHEP} {\bf 10} (2013)
  188},
\href{http://arxiv.org/abs/1307.3534}{{\tt arXiv:1307.3534 [hep-th]}}.

\bibitem{DHoker:2015sve}
E.~D'Hoker, M.~B. Green, and P.~Vanhove, ``{Proof of a modular relation between
  1-, 2- and 3-loop Feynman diagrams on a torus},''
  \href{http://dx.doi.org/10.1016/j.jnt.2017.07.022}{{\em J.\ Number Theory}
  (2018)  381},
\href{http://arxiv.org/abs/1509.00363}{{\tt arXiv:1509.00363 [hep-th]}}.

\bibitem{Basu:2015ayg}
A.~Basu, ``{Poisson equation for the Mercedes diagram in string theory at genus
  one},'' \href{http://dx.doi.org/10.1088/0264-9381/33/5/055005}{{\em Class.
  Quant. Grav.} {\bf 33} (2016) no.~5, 055005},
\href{http://arxiv.org/abs/1511.07455}{{\tt arXiv:1511.07455 [hep-th]}}.

\bibitem{Basu:2016xrt}
A.~Basu, ``{Poisson equation for the three loop ladder diagram in string theory
  at genus one},'' \href{http://dx.doi.org/10.1142/S0217751X16501694}{{\em Int.
  J. Mod. Phys.} {\bf A31} (2016) no.~32, 1650169},
\href{http://arxiv.org/abs/1606.02203}{{\tt arXiv:1606.02203 [hep-th]}}.

\bibitem{Basu:2016kli}
A.~Basu, ``{Proving relations between modular graph functions},''
  \href{http://dx.doi.org/10.1088/0264-9381/33/23/235011}{{\em Class. Quant.
  Grav.} {\bf 33} (2016) no.~23, 235011},
\href{http://arxiv.org/abs/1606.07084}{{\tt arXiv:1606.07084 [hep-th]}}.

\bibitem{Basu:2016mmk}
A.~Basu, ``{Simplifying the one loop five graviton amplitude in type IIB string
  theory},'' \href{http://dx.doi.org/10.1142/S0217751X17500749}{{\em Int. J.
  Mod. Phys.} {\bf A32} (2017) no.~14, 1750074},
\href{http://arxiv.org/abs/1608.02056}{{\tt arXiv:1608.02056 [hep-th]}}.

\bibitem{DHoker:2016quv}
E.~D'Hoker and J.~Kaidi, ``{Hierarchy of Modular Graph Identities},''
  \href{http://dx.doi.org/10.1007/JHEP11(2016)051}{{\em JHEP} {\bf 11} (2016)
  051},
\href{http://arxiv.org/abs/1608.04393}{{\tt arXiv:1608.04393 [hep-th]}}.

\bibitem{Kleinschmidt:2017ege}
A.~Kleinschmidt and V.~Verschinin, ``{Tetrahedral modular graph functions},''
  \href{http://dx.doi.org/10.1007/JHEP09(2017)155}{{\em JHEP} {\bf 09} (2017)
  155},
\href{http://arxiv.org/abs/1706.01889}{{\tt arXiv:1706.01889 [hep-th]}}.

\bibitem{Basu:2017nhs}
A.~Basu, ``{Low momentum expansion of one loop amplitudes in heterotic string
  theory},'' \href{http://dx.doi.org/10.1007/JHEP11(2017)139}{{\em JHEP} {\bf
  11} (2017)  139},
\href{http://arxiv.org/abs/1708.08409}{{\tt arXiv:1708.08409 [hep-th]}}.

\bibitem{Broedel:2018izr}
J.~Broedel, O.~Schlotterer, and F.~Zerbini, ``{From elliptic multiple zeta
  values to modular graph functions: open and closed strings at one loop},''
  \href{http://dx.doi.org/10.1007/JHEP01(2019)155}{{\em JHEP} {\bf 01} (2019)
  155},
\href{http://arxiv.org/abs/1803.00527}{{\tt arXiv:1803.00527 [hep-th]}}.

\bibitem{Ahlen:2018wng}
O.~Ahl{\'e}n and A.~Kleinschmidt, ``{$D^{6}R^{4}$ curvature corrections,
  modular graph functions and Poincar{\'e} series},''
  \href{http://dx.doi.org/10.1007/JHEP05(2018)194}{{\em JHEP} {\bf 05} (2018)
  194},
\href{http://arxiv.org/abs/1803.10250}{{\tt arXiv:1803.10250 [hep-th]}}.

\bibitem{Gerken:2018zcy}
J.~E. Gerken and J.~Kaidi, ``{Holomorphic subgraph reduction of higher-point
  modular graph forms},'' \href{http://dx.doi.org/10.1007/JHEP01(2019)131}{{\em
  JHEP} {\bf 01} (2019)  131},
\href{http://arxiv.org/abs/1809.05122}{{\tt arXiv:1809.05122 [hep-th]}}.

\bibitem{Gerken:2018jrq}
J.~E. Gerken, A.~Kleinschmidt, and O.~Schlotterer, ``{Heterotic-string
  amplitudes at one loop: modular graph forms and relations to open strings},''
  \href{http://dx.doi.org/10.1007/JHEP01(2019)052}{{\em JHEP} {\bf 01} (2019)
  052},
\href{http://arxiv.org/abs/1811.02548}{{\tt arXiv:1811.02548 [hep-th]}}.

\bibitem{DHoker:2019txf}
E.~D'Hoker and J.~Kaidi, ``{Modular graph functions and odd cuspidal functions.
  Fourier and Poincar{\'e} series},''
  \href{http://dx.doi.org/10.1007/JHEP04(2019)136}{{\em JHEP} {\bf 04} (2019)
  136},
\href{http://arxiv.org/abs/1902.04180}{{\tt arXiv:1902.04180 [hep-th]}}.

\bibitem{Dorigoni:2019yoq}
D.~Dorigoni and A.~Kleinschmidt, ``{Modular graph functions and asymptotic
  expansions of Poincar\'e series},''
  \href{http://dx.doi.org/10.4310/CNTP.2019.v13.n3.a3}{{\em Commun. Num. Theor.
  Phys.} {\bf 13} (2019) no.~3, 569--617},
\href{http://arxiv.org/abs/1903.09250}{{\tt arXiv:1903.09250 [hep-th]}}.

\bibitem{DHoker:2019xef}
E.~D'Hoker and M.~B. Green, ``{Absence of irreducible multiple zeta-values in
  melon modular graph functions},''
  \href{http://dx.doi.org/10.4310/CNTP.2020.v14.n2.a2}{{\em Commun. Num. Theor.
  Phys.} {\bf 14} (2020) no.~2, 315--324},
\href{http://arxiv.org/abs/1904.06603}{{\tt arXiv:1904.06603 [hep-th]}}.

\bibitem{DHoker:2019mib}
E.~D'Hoker, ``{Integral of two-loop modular graph functions},''
  \href{http://dx.doi.org/10.1007/JHEP06(2019)092}{{\em JHEP} {\bf 06} (2019)
  092},
\href{http://arxiv.org/abs/1905.06217}{{\tt arXiv:1905.06217 [hep-th]}}.

\bibitem{DHoker:2019blr}
E.~D'Hoker and M.~B. Green, ``{Exploring transcendentality in superstring
  amplitudes},'' \href{http://dx.doi.org/10.1007/JHEP07(2019)149}{{\em JHEP}
  {\bf 07} (2019)  149},
\href{http://arxiv.org/abs/1906.01652}{{\tt arXiv:1906.01652 [hep-th]}}.

\bibitem{Basu:2019idd}
A.~Basu, ``{Eigenvalue equation for the modular graph $C_{a,b,c,d}$},''
  \href{http://dx.doi.org/10.1007/JHEP07(2019)126}{{\em JHEP} {\bf 07} (2019)
  126},
\href{http://arxiv.org/abs/1906.02674}{{\tt arXiv:1906.02674 [hep-th]}}.

\bibitem{Gerken:2019cxz}
J.~E. Gerken, A.~Kleinschmidt, and O.~Schlotterer, ``{All-order differential
  equations for one-loop closed-string integrals and modular graph forms},''
  \href{http://dx.doi.org/10.1007/JHEP01(2020)064}{{\em JHEP} {\bf 01} (2020)
  064},
\href{http://arxiv.org/abs/1911.03476}{{\tt arXiv:1911.03476 [hep-th]}}.

\bibitem{Hohenegger:2019tii}
S.~Hohenegger, ``{From Little String Free Energies Towards Modular Graph
  Functions},'' \href{http://dx.doi.org/10.1007/JHEP03(2020)077}{{\em JHEP}
  {\bf 03} (2020)  077},
\href{http://arxiv.org/abs/1911.08172}{{\tt arXiv:1911.08172 [hep-th]}}.

\bibitem{Gerken:2020yii}
J.~E. Gerken, A.~Kleinschmidt, and O.~Schlotterer, ``{Generating series of all
  modular graph forms from iterated Eisenstein integrals},''
  \href{http://dx.doi.org/10.1007/JHEP07(2020)190}{{\em JHEP} {\bf 07} (2020)
  190}, \href{http://arxiv.org/abs/2004.05156}{{\tt arXiv:2004.05156
  [hep-th]}}.

\bibitem{Basu:2020kka}
A.~Basu, ``{Zero mode of the Fourier series of some modular graphs from
  Poincare series},''
  \href{http://dx.doi.org/10.1016/j.physletb.2020.135715}{{\em Phys. Lett. B}
  {\bf 809} (2020)  135715}, \href{http://arxiv.org/abs/2005.07793}{{\tt
  arXiv:2005.07793 [hep-th]}}.

\bibitem{Vanhove:2020qtt}
P.~Vanhove and F.~Zerbini, ``{Building blocks of closed and open string
  amplitudes},'' in {\em {MathemAmplitudes 2019: Intersection Theory and
  Feynman Integrals}}.
\newblock 7, 2020.
\newblock \href{http://arxiv.org/abs/2007.08981}{{\tt arXiv:2007.08981
  [hep-th]}}.

\bibitem{Basu:2020pey}
A.~Basu, ``{Poisson equations for elliptic modular graph functions},''
  \href{http://dx.doi.org/10.1016/j.physletb.2021.136086}{{\em Phys. Lett. B}
  {\bf 814} (2021)  136086}, \href{http://arxiv.org/abs/2009.02221}{{\tt
  arXiv:2009.02221 [hep-th]}}.

\bibitem{Basu:2020iok}
A.~Basu, ``{Relations between elliptic modular graphs},''
  \href{http://dx.doi.org/10.1007/JHEP12(2020)195}{{\em JHEP} {\bf 12} (2020)
  195}, \href{http://arxiv.org/abs/2010.08331}{{\tt arXiv:2010.08331
  [hep-th]}}. [Erratum: JHEP 03, 061 (2021)].

\bibitem{Hohenegger:2020slq}
S.~Hohenegger, ``{Diagrammatic Expansion of Non-Perturbative Little String Free
  Energies},'' \href{http://dx.doi.org/10.1007/JHEP04(2021)275}{{\em JHEP} {\bf
  04} (2021)  275}, \href{http://arxiv.org/abs/2011.06323}{{\tt
  arXiv:2011.06323 [hep-th]}}.

\bibitem{Dorigoni:2021jfr}
D.~Dorigoni, A.~Kleinschmidt, and O.~Schlotterer, ``{Poincar\'e series for
  modular graph forms at depth two. Part I. Seeds and Laplace systems},''
  \href{http://dx.doi.org/10.1007/JHEP01(2022)133}{{\em JHEP} {\bf 01} (2022)
  133}, \href{http://arxiv.org/abs/2109.05017}{{\tt arXiv:2109.05017
  [hep-th]}}.

\bibitem{Dorigoni:2021ngn}
D.~Dorigoni, A.~Kleinschmidt, and O.~Schlotterer, ``{Poincar\'e series for
  modular graph forms at depth two. Part II. Iterated integrals of cusp
  forms},'' \href{http://dx.doi.org/10.1007/JHEP01(2022)134}{{\em JHEP} {\bf
  01} (2022)  134}, \href{http://arxiv.org/abs/2109.05018}{{\tt
  arXiv:2109.05018 [hep-th]}}.

\bibitem{Brown:mmv}
F.~Brown, ``{Multiple modular values and the relative completion of the
  fundamental group of ${\cal M}_{1,1}$},''
  \href{http://arxiv.org/abs/1407.5167}{{\tt arXiv:1407.5167 [math.NT]}}.

\bibitem{Zerbini:2015rss}
F.~Zerbini, ``{Single-valued multiple zeta values in genus 1 superstring
  amplitudes},'' \href{http://dx.doi.org/10.4310/CNTP.2016.v10.n4.a2}{{\em
  Commun. Num. Theor. Phys.} {\bf 10} (2016) no.~4, 703--737},
\href{http://arxiv.org/abs/1512.05689}{{\tt arXiv:1512.05689 [hep-th]}}.

\bibitem{Brown:I}
F.~Brown, ``A class of non-holomorphic modular forms {I},''
  \href{http://dx.doi.org/10.1007/s40687-018-0130-8}{{\em Res. Math. Sci.} {\bf
  5} (2018) no.~1, Paper No. 7, 40},
  \href{http://arxiv.org/abs/1707.01230}{{\tt arXiv:1707.01230 [math.NT]}}.

\bibitem{Brown:II}
F.~Brown, ``A class of nonholomorphic modular forms {II}: {E}quivariant
  iterated {E}isenstein integrals,''
  \href{http://dx.doi.org/10.1017/fms.2020.24}{{\em Forum Math. Sigma} {\bf 8}
  (2020)  Paper No. e31, 62}, \href{http://arxiv.org/abs/1708.03354}{{\tt
  arXiv:1708.03354 [math.NT]}}.

\bibitem{DHoker:2017zhq}
E.~D'Hoker and W.~Duke, ``Fourier series of modular graph functions,''
  \href{http://dx.doi.org/10.1016/j.jnt.2018.04.012}{{\em J. Number Theory}
  {\bf 192} (2018)  1--36}, \href{http://arxiv.org/abs/1708.07998}{{\tt
  arXiv:1708.07998 [math.NT]}}.

\bibitem{Zerbini:2018sox}
F.~Zerbini, {\em {Elliptic multiple zeta values, modular graph functions and
  genus 1 superstring scattering amplitudes}}.
\newblock PhD thesis, Bonn U., 2017.
\newblock
\href{http://arxiv.org/abs/1804.07989}{{\tt arXiv:1804.07989 [math-ph]}}.
\newblock

\bibitem{Zerbini:2018hgs}
F.~Zerbini, ``{Modular and holomorphic graph function from superstring
  amplitudes},'' in {\em {KMPB Conference: Elliptic Integrals, Elliptic
  Functions and Modular Forms in Quantum Field Theory Zeuthen, Germany, October
  23-26, 2017}}.
\newblock 2018.
\newblock
\href{http://arxiv.org/abs/1807.04506}{{\tt arXiv:1807.04506 [math-ph]}}.
\newblock

\bibitem{Zagier:2019eus}
D.~Zagier and F.~Zerbini, ``{Genus-zero and genus-one string amplitudes and
  special multiple zeta values},''
  \href{http://dx.doi.org/10.4310/CNTP.2020.v14.n2.a4}{{\em Commun. Num. Theor.
  Phys.} {\bf 14} (2020) no.~2, 413--452},
\href{http://arxiv.org/abs/1906.12339}{{\tt arXiv:1906.12339 [math.NT]}}.

\bibitem{Berg:2019jhh}
M.~Berg, K.~Bringmann, and T.~Gannon, ``{Massive deformations of Maass forms
  and Jacobi forms},''
  \href{http://dx.doi.org/10.4310/CNTP.2021.v15.n3.a4}{{\em Commun. Num. Theor.
  Phys.} {\bf 15} (2021) no.~3, 575--603},
  \href{http://arxiv.org/abs/1910.02745}{{\tt arXiv:1910.02745 [math.NT]}}.

\bibitem{Drewitt:2021}
J.~Drewitt, ``Laplace-eigenvalue equations for length three modular iterated
  integrals,'' \href{http://dx.doi.org/10.1016/j.jnt.2021.11.005}{{\em J.
  Number Theory} {\bf 239} (2022)  78--112},
  \href{http://arxiv.org/abs/2104.09916}{{\tt arXiv:2104.09916 [math.NT]}}.

\bibitem{Gerken:review}
J.~E. Gerken, \href{http://dx.doi.org/http://dx.doi.org/10.18452/21829}{{\em
  {Modular Graph Forms and Scattering Amplitudes in String Theory}}}.
\newblock PhD thesis, {Humboldt-Universität zu Berlin}, 2020.
\newblock \href{http://arxiv.org/abs/2011.08647}{{\tt arXiv:2011.08647
  [hep-th]}}.

\bibitem{Gerken:2020aju}
J.~E. Gerken, ``{Basis Decompositions and a Mathematica Package for Modular
  Graph Forms},'' \href{http://dx.doi.org/10.1088/1751-8121/abbdf2}{{\em J.
  Phys. A} {\bf 54} (2021) no.~19, 195401},
  \href{http://arxiv.org/abs/2007.05476}{{\tt arXiv:2007.05476 [hep-th]}}.

\bibitem{DHoker:2022dxx}
E.~D'Hoker and J.~Kaidi, ``{Lectures on modular forms and strings},''
  \href{http://arxiv.org/abs/2208.07242}{{\tt arXiv:2208.07242 [hep-th]}}.

\bibitem{Zagier:notes}
D.~Zagier, ``Evaluation of lattice sums,''. unpublished notes.

\bibitem{Iwaniec:2002}
H.~Iwaniec, \href{http://dx.doi.org/10.1090/gsm/053}{{\em {Spectral methods of
  automorphic forms}}}, vol.~53 of {\em Graduate Studies in Mathematics}.
\newblock American Mathematical Society, Providence, RI; Revista Matem\'{a}tica
  Iberoamericana, Madrid, second~ed., 2002.

\bibitem{Fleig:2015vky}
P.~Fleig, H.~P.~A. Gustafsson, A.~Kleinschmidt, and D.~Persson,
  \href{http://dx.doi.org/10.1017/9781316995860}{{\em Eisenstein series and
  automorphic representations, with applications in string theory}}, vol.~176
  of {\em Cambridge Studies in Advanced Mathematics}.
\newblock Cambridge University Press, Cambridge, 2018.
\newblock \href{http://arxiv.org/abs/1511.04265}{{\tt arXiv:1511.04265
  [math.NT]}}.

\bibitem{Green:2014yxa}
M.~B. Green, S.~D. Miller, and P.~Vanhove, ``{$SL(2, \mathbb{Z})$-invariance
  and D-instanton contributions to the $D^6 R^4$ interaction},''
  \href{http://dx.doi.org/10.4310/CNTP.2015.v9.n2.a3}{{\em Commun. Num. Theor.
  Phys.} {\bf 09} (2015)  307--344},
\href{http://arxiv.org/abs/1404.2192}{{\tt arXiv:1404.2192 [hep-th]}}.

\bibitem{Dorigoni:2014hea}
D.~Dorigoni, ``{An Introduction to Resurgence, Trans-Series and Alien
  Calculus},'' \href{http://dx.doi.org/10.1016/j.aop.2019.167914}{{\em Annals
  Phys.} {\bf 409} (2019)  167914}, \href{http://arxiv.org/abs/1411.3585}{{\tt
  arXiv:1411.3585 [hep-th]}}.

\bibitem{Dorigoni:2020oon}
D.~Dorigoni and A.~Kleinschmidt, ``{Resurgent expansion of Lambert series and
  iterated Eisenstein integrals},''
  \href{http://dx.doi.org/10.4310/CNTP.2021.v15.n1.a1}{{\em Commun. Num. Theor.
  Phys.} {\bf 15} (2021) no.~1, 1--57},
  \href{http://arxiv.org/abs/2001.11035}{{\tt arXiv:2001.11035 [hep-th]}}.

\bibitem{Dunne:2016jsr}
G.~V. Dunne and M.~Unsal, ``{Deconstructing zero: resurgence, supersymmetry and
  complex saddles},'' \href{http://dx.doi.org/10.1007/JHEP12(2016)002}{{\em
  JHEP} {\bf 12} (2016)  002}, \href{http://arxiv.org/abs/1609.05770}{{\tt
  arXiv:1609.05770 [hep-th]}}.

\bibitem{Kozcaz:2016wvy}
C.~Koz\c{c}az, T.~Sulejmanpasic, Y.~Tanizaki, and M.~\"Unsal, ``{Cheshire Cat
  resurgence, Self-resurgence and Quasi-Exact Solvable Systems},''
  \href{http://dx.doi.org/10.1007/s00220-018-3281-y}{{\em Commun. Math. Phys.}
  {\bf 364} (2018) no.~3, 835--878},
  \href{http://arxiv.org/abs/1609.06198}{{\tt arXiv:1609.06198 [hep-th]}}.

\bibitem{Dorigoni:2017smz}
D.~Dorigoni and P.~Glass, ``{The grin of Cheshire cat resurgence from
  supersymmetric localization},''
  \href{http://dx.doi.org/10.21468/SciPostPhys.4.2.012}{{\em SciPost Phys.}
  {\bf 4} (2018) no.~2, 012}, \href{http://arxiv.org/abs/1711.04802}{{\tt
  arXiv:1711.04802 [hep-th]}}.

\bibitem{Dorigoni:2019kux}
D.~Dorigoni and P.~Glass, ``{Picard-Lefschetz decomposition and Cheshire Cat
  resurgence in 3D $ \mathcal{N} $ = 2 field theories},''
  \href{http://dx.doi.org/10.1007/JHEP12(2019)085}{{\em JHEP} {\bf 12} (2019)
  085}, \href{http://arxiv.org/abs/1909.05262}{{\tt arXiv:1909.05262
  [hep-th]}}.

\bibitem{ApostolTomM1976MfaD}
T.~M. Apostol, \href{http://dx.doi.org/10.1007/978-1-4612-0999-7}{{\em Modular
  functions and Dirichlet series in number theory.}}
\newblock Graduate texts in mathematics; 41. Springer-Verlag, New York, 1976.

\bibitem{Brown:2013gia}
F.~Brown, ``{Single-valued Motivic Periods and Multiple Zeta Values},''
  \href{http://dx.doi.org/10.1017/fms.2014.18}{{\em SIGMA} {\bf 2} (2014)
  e25},
\href{http://arxiv.org/abs/1309.5309}{{\tt arXiv:1309.5309 [math.NT]}}.

\bibitem{DHoker:2021ous}
E.~D'Hoker and N.~Geiser, ``{Integrating three-loop modular graph functions and
  transcendentality of string amplitudes},''
  \href{http://dx.doi.org/10.1007/JHEP02(2022)019}{{\em JHEP} {\bf 02} (2022)
  019}, \href{http://arxiv.org/abs/2110.06237}{{\tt arXiv:2110.06237
  [hep-th]}}.

\bibitem{Green:2008bf}
M.~B. Green, J.~G. Russo, and P.~Vanhove, ``{Modular properties of two-loop
  maximal supergravity and connections with string theory},''
  \href{http://dx.doi.org/10.1088/1126-6708/2008/07/126}{{\em JHEP} {\bf 07}
  (2008)  126},
\href{http://arxiv.org/abs/0807.0389}{{\tt arXiv:0807.0389 [hep-th]}}.

\bibitem{Broedel:2015hia}
J.~Broedel, N.~Matthes, and O.~Schlotterer, ``{Relations between elliptic
  multiple zeta values and a special derivation algebra},''
  \href{http://dx.doi.org/10.1088/1751-8113/49/15/155203}{{\em J. Phys.} {\bf
  A49} (2016) no.~15, 155203},
\href{http://arxiv.org/abs/1507.02254}{{\tt arXiv:1507.02254 [hep-th]}}.

\bibitem{Arutyunov:2016etw}
G.~Arutyunov, D.~Dorigoni, and S.~Savin, ``{Resurgence of the dressing phase
  for AdS$_{5} \times$ S$^{5}$},''
  \href{http://dx.doi.org/10.1007/JHEP01(2017)055}{{\em JHEP} {\bf 01} (2017)
  055}, \href{http://arxiv.org/abs/1608.03797}{{\tt arXiv:1608.03797
  [hep-th]}}.

\bibitem{Klinger:2018}
K.~Klinger-Logan, ``Differential equations in automorphic forms,''
  \href{http://dx.doi.org/10.4310/CNTP.2018.v12.n4.a4}{{\em Commun. Number
  Theory Phys.} {\bf 12} (2018) no.~4, 767--827},
  \href{http://arxiv.org/abs/1801.00838}{{\tt arXiv:1801.00838 [math.NT]}}.

\bibitem{Collier:2022emf}
S.~Collier and E.~Perlmutter, ``{Harnessing S-duality in $ \mathcal{N} $ = 4
  SYM \& supergravity as SL(2, \ensuremath{\mathbb{Z}})-averaged strings},''
  \href{http://dx.doi.org/10.1007/JHEP08(2022)195}{{\em JHEP} {\bf 08} (2022)
  195}, \href{http://arxiv.org/abs/2201.05093}{{\tt arXiv:2201.05093
  [hep-th]}}.

\bibitem{Green:2005ba}
M.~B. Green and P.~Vanhove, ``{Duality and higher derivative terms in M
  theory},'' \href{http://dx.doi.org/10.1088/1126-6708/2006/01/093}{{\em JHEP}
  {\bf 01} (2006)  093},
\href{http://arxiv.org/abs/hep-th/0510027}{{\tt arXiv:hep-th/0510027
  [hep-th]}}.

\bibitem{Green:1997tv}
M.~B. Green and M.~Gutperle, ``{Effects of D instantons},''
  \href{http://dx.doi.org/10.1016/S0550-3213(97)00269-1}{{\em Nucl. Phys.} {\bf
  B498} (1997)  195--227},
\href{http://arxiv.org/abs/hep-th/9701093}{{\tt arXiv:hep-th/9701093
  [hep-th]}}.

\bibitem{Green:1998by}
M.~B. Green and S.~Sethi, ``{Supersymmetry constraints on type IIB
  supergravity},'' \href{http://dx.doi.org/10.1103/PhysRevD.59.046006}{{\em
  Phys. Rev. D} {\bf 59} (1999)  046006},
  \href{http://arxiv.org/abs/hep-th/9808061}{{\tt arXiv:hep-th/9808061}}.

\bibitem{Pioline:2015yea}
B.~Pioline, ``{D$^{6}$R$^{4}$ amplitudes in various dimensions},''
  \href{http://dx.doi.org/10.1007/JHEP04(2015)057}{{\em JHEP} {\bf 04} (2015)
  057}, \href{http://arxiv.org/abs/1502.03377}{{\tt arXiv:1502.03377
  [hep-th]}}.

\bibitem{Bossard:2015uga}
G.~Bossard and V.~Verschinin, ``{The two \ensuremath{\nabla}$^{6}$ R$^{4}$ type
  invariants and their higher order generalisation},''
  \href{http://dx.doi.org/10.1007/JHEP07(2015)154}{{\em JHEP} {\bf 07} (2015)
  154}, \href{http://arxiv.org/abs/1503.04230}{{\tt arXiv:1503.04230
  [hep-th]}}.

\bibitem{Bossard:2020xod}
G.~Bossard, A.~Kleinschmidt, and B.~Pioline, ``{1/8-BPS Couplings and
  Exceptional Automorphic Functions},''
  \href{http://dx.doi.org/10.21468/SciPostPhys.8.4.054}{{\em SciPost Phys.}
  {\bf 8} (2020) no.~4, 054}, \href{http://arxiv.org/abs/2001.05562}{{\tt
  arXiv:2001.05562 [hep-th]}}.

\bibitem{Chester:2020vyz}
S.~M. Chester, M.~B. Green, S.~S. Pufu, Y.~Wang, and C.~Wen, ``{New modular
  invariants in $ \mathcal{N} $ = 4 Super-Yang-Mills theory},''
  \href{http://dx.doi.org/10.1007/JHEP04(2021)212}{{\em JHEP} {\bf 04} (2021)
  212}, \href{http://arxiv.org/abs/2008.02713}{{\tt arXiv:2008.02713
  [hep-th]}}.

\bibitem{Minprogress}
K.~Klinger-Logan, S.~D. Miller, and D.~Radchenko, ``{The $D^6R^4$ interaction
  as a Poincar\'e series, and a related shifted convolution sum},'' {\em work
  in progress}  .

\bibitem{FKinprogress}
K.~Fedosova and K.~Klinger-Logan, ``{Whittaker Fourier type solutions to
  differential equations arising from string theory},'' {\em work in progress}
  .

\bibitem{Binder:2019jwn}
D.~J. Binder, S.~M. Chester, S.~S. Pufu, and Y.~Wang, ``{$ \mathcal{N} $ = 4
  Super-Yang-Mills correlators at strong coupling from string theory and
  localization},'' \href{http://dx.doi.org/10.1007/JHEP12(2019)119}{{\em JHEP}
  {\bf 12} (2019)  119}, \href{http://arxiv.org/abs/1902.06263}{{\tt
  arXiv:1902.06263 [hep-th]}}.

\bibitem{Binder:2019mpb}
D.~J. Binder, S.~M. Chester, and S.~S. Pufu, ``{AdS$_{4}$/CFT$_{3}$ from weak
  to strong string coupling},''
  \href{http://dx.doi.org/10.1007/JHEP01(2020)034}{{\em JHEP} {\bf 01} (2020)
  034}, \href{http://arxiv.org/abs/1906.07195}{{\tt arXiv:1906.07195
  [hep-th]}}.

\bibitem{Chester:2019pvm}
S.~M. Chester, ``{Genus-2 holographic correlator on AdS$_{5} \times $ S$^{5}$
  from localization},'' \href{http://dx.doi.org/10.1007/JHEP04(2020)193}{{\em
  JHEP} {\bf 04} (2020)  193}, \href{http://arxiv.org/abs/1908.05247}{{\tt
  arXiv:1908.05247 [hep-th]}}.

\bibitem{Chester:2019jas}
S.~M. Chester, M.~B. Green, S.~S. Pufu, Y.~Wang, and C.~Wen, ``{Modular
  Invariance in Superstring Theory From ${\cal N} = 4$ Super-Yang-Mills},''
  \href{http://dx.doi.org/10.1007/JHEP11(2020)016}{{\em JHEP} {\bf 11} (2020)
  016}, \href{http://arxiv.org/abs/1912.13365}{{\tt arXiv:1912.13365
  [hep-th]}}.

\bibitem{Green:2020eyj}
M.~B. Green and C.~Wen, ``{Maximal U(1)$_{Y}$-violating n-point correlators in
  $ \mathcal{N} $ = 4 super-Yang-Mills theory},''
  \href{http://dx.doi.org/10.1007/JHEP02(2021)042}{{\em JHEP} {\bf 02} (2021)
  042}, \href{http://arxiv.org/abs/2009.01211}{{\tt arXiv:2009.01211
  [hep-th]}}.

\bibitem{Dorigoni:2021bvj}
D.~Dorigoni, M.~B. Green, and C.~Wen, ``{Novel Representation of an Integrated
  Correlator in $\mathcal N$ = 4 Supersymmetric Yang-Mills Theory},''
  \href{http://dx.doi.org/10.1103/PhysRevLett.126.161601}{{\em Phys. Rev.
  Lett.} {\bf 126} (2021) no.~16, 161601},
  \href{http://arxiv.org/abs/2102.08305}{{\tt arXiv:2102.08305 [hep-th]}}.

\bibitem{Dorigoni:2021guq}
D.~Dorigoni, M.~B. Green, and C.~Wen, ``{Exact properties of an integrated
  correlator in $ \mathcal{N} $ = 4 SU(N) SYM},''
  \href{http://dx.doi.org/10.1007/JHEP05(2021)089}{{\em JHEP} {\bf 05} (2021)
  089}, \href{http://arxiv.org/abs/2102.09537}{{\tt arXiv:2102.09537
  [hep-th]}}.

\bibitem{Okuda:2010ym}
T.~Okuda and J.~Penedones, ``{String scattering in flat space and a scaling
  limit of Yang-Mills correlators},''
  \href{http://dx.doi.org/10.1103/PhysRevD.83.086001}{{\em Phys. Rev. D} {\bf
  83} (2011)  086001}, \href{http://arxiv.org/abs/1002.2641}{{\tt
  arXiv:1002.2641 [hep-th]}}.

\bibitem{Penedones:2010ue}
J.~Penedones, ``{Writing CFT correlation functions as AdS scattering
  amplitudes},'' \href{http://dx.doi.org/10.1007/JHEP03(2011)025}{{\em JHEP}
  {\bf 03} (2011)  025}, \href{http://arxiv.org/abs/1011.1485}{{\tt
  arXiv:1011.1485 [hep-th]}}.

\bibitem{Alday:2018pdi}
L.~F. Alday, A.~Bissi, and E.~Perlmutter, ``{Genus-One String Amplitudes from
  Conformal Field Theory},''
  \href{http://dx.doi.org/10.1007/JHEP06(2019)010}{{\em JHEP} {\bf 06} (2019)
  010},
\href{http://arxiv.org/abs/1809.10670}{{\tt arXiv:1809.10670 [hep-th]}}.

\bibitem{Alday:2018kkw}
L.~F. Alday, ``{On genus-one string amplitudes on $AdS_5 \times S^5$},''
  \href{http://dx.doi.org/10.1007/JHEP04(2021)005}{{\em JHEP} {\bf 04} (2021)
  005}, \href{http://arxiv.org/abs/1812.11783}{{\tt arXiv:1812.11783
  [hep-th]}}.

\bibitem{Drummond:2019odu}
J.~Drummond, D.~Nandan, H.~Paul, and K.~Rigatos, ``{String corrections to AdS
  amplitudes and the double-trace spectrum of $ \mathcal{N} $ = 4 SYM},''
  \href{http://dx.doi.org/10.1007/JHEP12(2019)173}{{\em JHEP} {\bf 12} (2019)
  173}, \href{http://arxiv.org/abs/1907.00992}{{\tt arXiv:1907.00992
  [hep-th]}}.

\bibitem{Aprile:2019rep}
F.~Aprile, J.~Drummond, P.~Heslop, and H.~Paul, ``{One-loop amplitudes in
  AdS$_{5} \times$ S$^{5}$ supergravity from $ \mathcal{N} $ = 4 SYM at strong
  coupling},'' \href{http://dx.doi.org/10.1007/JHEP03(2020)190}{{\em JHEP} {\bf
  03} (2020)  190}, \href{http://arxiv.org/abs/1912.01047}{{\tt
  arXiv:1912.01047 [hep-th]}}.

\bibitem{Drummond:2019hel}
J.~M. Drummond and H.~Paul, ``{One-loop string corrections to AdS amplitudes
  from CFT},'' \href{http://dx.doi.org/10.1007/JHEP03(2021)038}{{\em JHEP} {\bf
  03} (2021)  038}, \href{http://arxiv.org/abs/1912.07632}{{\tt
  arXiv:1912.07632 [hep-th]}}.

\bibitem{Bissi:2020wtv}
A.~Bissi, G.~Fardelli, and A.~Georgoudis, ``{Towards all loop supergravity
  amplitudes on AdS5$\times$S5},''
  \href{http://dx.doi.org/10.1103/PhysRevD.104.L041901}{{\em Phys. Rev. D} {\bf
  104} (2021) no.~4, L041901}, \href{http://arxiv.org/abs/2002.04604}{{\tt
  arXiv:2002.04604 [hep-th]}}.

\bibitem{Bissi:2020woe}
A.~Bissi, G.~Fardelli, and A.~Georgoudis, ``{All loop structures in
  supergravity amplitudes on AdS5$\times$S5 from CFT},''
  \href{http://dx.doi.org/10.1088/1751-8121/ac0ebf}{{\em J. Phys. A} {\bf 54}
  (2021) no.~32, 324002}, \href{http://arxiv.org/abs/2010.12557}{{\tt
  arXiv:2010.12557 [hep-th]}}.

\bibitem{Dorigoni:2021rdo}
D.~Dorigoni, M.~B. Green, and C.~Wen, ``{Exact expressions for $n$-point
  maximal $U(1)_Y$-violating integrated correlators in $SU(N)\,\mathcal{N}=4$
  SYM},'' \href{http://dx.doi.org/10.1007/JHEP11(2021)132}{{\em JHEP} {\bf 11}
  (2021)  132}, \href{http://arxiv.org/abs/2109.08086}{{\tt arXiv:2109.08086
  [hep-th]}}.

\bibitem{Dorigoni:2022zcr}
D.~Dorigoni, M.~B. Green, and C.~Wen, ``{Exact results for duality-covariant
  integrated correlators in $\mathcal{N}=4$ SYM with general classical gauge
  groups},'' \href{http://arxiv.org/abs/2202.05784}{{\tt arXiv:2202.05784
  [hep-th]}}.

\bibitem{Dorigoni:2022iem}
D.~Dorigoni, M.~B. Green, and C.~Wen, ``{The SAGEX Review on Scattering
  Amplitudes, Chapter 10: Modular covariance of type IIB string amplitudes and
  their $\mathcal{N}=4$ supersymmetric Yang-Mills duals},''
  \href{http://arxiv.org/abs/2203.13021}{{\tt arXiv:2203.13021 [hep-th]}}.

\bibitem{Hatsdua:2022enx}
Y.~Hatsuda and K.~Okuyama, ``{Large $N$ expansion of an integrated correlator
  in $\mathcal{N}=4$ SYM},'' \href{http://arxiv.org/abs/2208.01891}{{\tt
  arXiv:2208.01891 [hep-th]}}.

\bibitem{DHoker:2020hlp}
E.~D'Hoker, A.~Kleinschmidt, and O.~Schlotterer, ``{Elliptic modular graph
  forms. Part I. Identities and generating series},''
  \href{http://dx.doi.org/10.1007/JHEP03(2021)151}{{\em JHEP} {\bf 03} (2021)
  151}, \href{http://arxiv.org/abs/2012.09198}{{\tt arXiv:2012.09198
  [hep-th]}}.

\bibitem{Hidding:2022vjf}
M.~Hidding, O.~Schlotterer, and B.~Verbeek, ``{Elliptic modular graph forms II:
  Iterated integrals},'' \href{http://arxiv.org/abs/2208.11116}{{\tt
  arXiv:2208.11116 [hep-th]}}.

\bibitem{Costin2008}
O.~Costin and S.~Garoufalidis, ``{Resurgence of the Euler-MacLaurin summation
  formula},'' {\em {Annales de l’institut Fourier}} {\bf {58}} ({2008})
  no.~{3}, {893--914}. \url{http://eudml.org/doc/10338}.

\bibitem{Crew:2022qoe}
S.~Crew and P.~H. Trinh, ``{Resurgent aspects of applied exponential
  asymptotics},'' \href{http://arxiv.org/abs/2208.07290}{{\tt arXiv:2208.07290
  [math.CA]}}.

\end{thebibliography}\endgroup
\bibliographystyle{utphys}

\end{document}